\documentclass[fleqn,usenatbib]{mnras}

\usepackage[T1]{fontenc}
\usepackage{ae,aecompl}

\usepackage{graphicx}	% Including figure files
\usepackage{amsmath}	% Advanced maths commands
\usepackage{lscape}
\usepackage{longtable}
\usepackage{supertabular, xcolor, multicol}
\usepackage{threeparttable}
\usepackage{threeparttablex}
\usepackage{float}
\usepackage[normalem]{ulem}

\newcommand{\blue}{\textcolour{\blue}}
\newcommand{\Ha}{\ifmmode \text{H}\alpha \else H$\alpha$\fi\xspace} % added 20200221
\newcommand{\nii}{\ifmmode [\text{N}\,\textsc{ii}] \else [N~{\scshape ii}]\fi\xspace} % added 20200221
\title[Panchromatic view of Hydra Cluster]{An environmental dependence of the physical and structural properties in the Hydra Cluster galaxies}

\author[C. Lima-Dias et al.]{
Ciria Lima-Dias$^{1}$\thanks{E-mail: clima@userena.cl},
Antonela Monachesi$^{1,2}$,
Sergio Torres-Flores$^{1}$,
Arianna Cortesi$^{3,4}$, \newauthor
Daniel Hern\'{a}ndez-Lang$^{5}$,
Carlos Eduardo Barbosa$^{3,6}$,
Claudia Mendes de Oliveira$^{3}$,\newauthor
Daniela Olave-Rojas$^{7}$,
Diego Pallero$^{1}$,
Laura Sampedro$^{3}$,
Alberto Molino$^{3}$,\newauthor 
Fabio R. Herpich$^{3}$,
Yara L. Jaff\'e$^{8}$,
Ricardo Amor\'in$^{1,2}$,
Ana L. Chies-Santos$^{9}$,\newauthor
Paola Dimauro$^{10}$,
Eduardo Telles$^{10}$,
Paulo A. A. Lopes$^{4}$, 
Alvaro Alvarez-Candal$^{10,11}$,\newauthor
Fabricio Ferrari$^{12}$,
Antonio Kanaan$^{13}$,
Tiago Ribeiro$^{14}$,
William Schoenell$^{15}$
\\
$^{1}$ Departamento de Astronom\'ia, Universidad de La Serena, Av. Cisternas 1200, La Serena, Chile\\
$^{2}$ Instituto de Investigaci\'on Multidisciplinar en Ciencia y Tecnolog\'ia, Universidad de La Serena, Ra\'ul Bitr\'an 1305, La Serena, Chile\\
$^{3}$ Instituto de Astronomia, Geof\'isica e Ci\^encias Atmosf\'ericas da Universidade de S\~ao Paulo, Cidade Universit\'aria, CEP:05508-900, \\ S\~ao Paulo, SP, Brazil\\
$^{4}$ Observat\'orio do Valongo, Universidade Federal do Rio de
Janeiro, Ladeira Pedro Ant\^onio 43,
Rio de Janeiro, RJ, 20080-090, Brazil\\
$^{5}$ Faculty of Physics, Ludwig-Maximilians-Universit\"{a}t, Scheinerstr.\ 1, 81679 Munich, Germany\\
$^{6}$Steward Observatory, University of Arizona, 933 N Cherry Ave, Tucson, AZ 85719, USA\\
$^{7}$ Departamento de Tecnolog\'ias Industriales, Universidad de Talca, Los Niches km 1, Curic\'o, Chile\\
$^{8}$ Instituto de F\'isica y Astronom\'ia, Universidad de Valpara\'iso, Avda. Gran Breta\~na 1111 Valpara\'iso, Chile\\ 
$^{9}$ Departamento de Astronomia, Instituto de F\'isica, Universidade Federal do Rio Grande do Sul (UFRGS), Av. Bento Gon\c{c}alves 9500, \\ Porto Alegre, RS, Brazil \\
$^{10}$ Observat\'orio Nacional / MCTIC, Rua General Jos\'e Cristino 77, Rio de Janeiro, RJ, 20921-400, Brazil\\ 
$^{11}$ Instituto Universitario de F\'isica Aplicada a las Ciencias y las Tecnolog\'ias, Universidad de Alicante, San Vicent del Raspeig, E03080, \\ Alicante, Spain\\
$^{12}$ Instituto de Matem\'atica Estat\'istica e F\'isica, Universidade
Federal do Rio Grande, Rio Grande, RS, 96201-900, Brazil\\
$^{13}$ Departamento de F\'isica, Centro de Ciencias F\'isicas e Matem\'aticas, Universidade Federal de Santa Catarina, Florian\'opolis, SC, \\ 88040-900, Brazil \\
$^{14}$ LSST Project Office, 950 N. Cherry Ave., Tucson, AZ 85719, USA \\
$^{15}$ GMTO Corporation 465 N. Halstead Street, Suite 250 Pasadena, CA 91107
}

\date{Accepted XXX. Received YYY; in original form ZZZ}

\pubyear{2019}

\begin{document}
\label{firstpage}
\pagerange{\pageref{firstpage}--\pageref{lastpage}}
\maketitle

\begin{abstract}
 The nearby Hydra Cluster ($\sim$50 Mpc) is an ideal laboratory to understand, in detail, the influence of the environment on the morphology and quenching of galaxies in dense environments. We study the Hydra cluster galaxies in the inner regions ($1R_{200}$) of the cluster using data from the Southern Photometric Local Universe Survey (S-PLUS), which uses 12 narrow and broad band filters in the visible region of the spectrum. We analyse structural (S\'ersic index, effective radius) and physical (colours, stellar masses and star formation rates) properties. Based on this analysis, we find that $\sim$88 percent of the Hydra cluster galaxies are quenched. Using the Dressler-Schectman test approach, we also find that the cluster shows possible substructures. Our analysis of the phase-space diagram together with {\tt DBSCAN} algorithm indicates that Hydra shows an additional substructure that appears to be in front of the cluster centre, which is still falling into it. Our results, thus, suggest that the Hydra Cluster might not be relaxed. We analyse the  median S\'ersic index as a function of wavelength and find that for red ($(u-r)\geq$2.3) and early-type galaxies it displays a slight increase towards redder filters (13 and 18 percent, for red and early-type respectively) whereas for blue+green ($(u-r)$<2.3) galaxies it remains constant. Late-type galaxies show a small decrease of the median S\'ersic index toward redder filters. Also, the S\'ersic index of galaxies, and thus their structural properties, do not significantly vary as a function of clustercentric distance and density within the cluster; and this is the case regardless of the filter.

\end{abstract}

\begin{keywords}
galaxies: clusters: general -- galaxies: fundamental parameters -- galaxies: structure
\end{keywords}

%%%%%%%%%%%%%%%%%%%%%%%%%%%%%%%%%%%%%%%%%%%%%%%%%%

%%%%%%%%%%%%%%%%% BODY OF PAPER %%%%%%%%%%%%%%%%%%

\section{Introduction}

One of the many remarkable and still open questions in extragalactic astronomy is: How do galaxies evolve with/within environment? Several studies have addressed this issue, generally based on pioneer works, which determined a relation between the environment in which galaxies reside and their physical properties \citep[]{Oemler74,Davis76,Butcher78,Dressler1980,Postman84}.
In particular \citet{Dressler1980} detected an increase in the fraction of elliptical and S0 galaxies (early-type galaxies, or ETGs) as a function of increasing the environmental density, while the opposite is observed for spiral galaxies (late-type galaxies, or LTGs) \citep{Gunn1972,Whitmore1991ApJ,Poggianti01,Boselli05,Fasano2015}. 
This indicates that the environment is playing a crucial role on the morphology and stellar production of galaxies and is one of the main mechanisms often associated to the star-formation (SF) quenching on the Local Universe.  In addition, internal process such as mass quenching, which is driven by gas outflows produced by stellar winds and supernovae feedback \citep[]{Larson74,Dekel86,Efstathiou00,Cantalupo10,Peng2010ApJ} or due to active galactic nuclei (AGN) activity \citep[]{Croton06,Fabian2012ARA&A,Cicone14}) also has its influence. In the case of AGN activity, will be specially relevant for massive galaxies ($M_\star 10^{10}M_{\odot}$) \citep{Peng2010ApJ,Cora2019}.

Environmental effects are drastically enhanced in clusters due to the tremendous gravitational potential. Fast and aggressive encounters between satellite galaxies become recurring. This process, known as harassment, is capable to destroy the disks of the satellites affected, becoming specially relevant in the inner parts of clusters \citep{Moore98,Moore99,Duc08,Smith15}. On the other hand, interaction between the galaxy and the intracluster medium can strip the gas component of galaxies. This phenomena is called ram-pressure or strangulation \citep{Gunn1972,Abadi1999,Quilis00,Balogh2000,Vollmer01,Jaffe2015,Peng2015}, depending the strength of the stripping.
Within a cluster, the galaxies inhabiting the core are usually the most massive ones, and most of them being quenched at early times. LTGs within clusters, are often found among the satellites at the outskirts \citep{Dressler97,Fasano00,Postman05,Desai07}. In general LTGs are the less massive components and are strongly susceptible to the influence of the processes aforementioned.

More recently using data from  the Sloan Digital Sky Survey (SDSS; \citealt{Abazajian2009}), \citet{Liu2019} compared the fractions of ETGs and LTGs, and the fraction of main sequence galaxies (MSG) and quenched galaxies (QG) relative to different environments (voids, sheets, filaments, and clusters), finding that the star-forming properties of galaxies changed more dramatically than their structural properties. This means that a galaxy will stop forming stars before any morphological change can take place. 
They also found that the morphological transformation and quenching for low mass galaxies can be two independent processes, suggesting that the interruption of the SF is determined by the halo mass while the morphological transformation is more correlated with the stellar mass.

While to-date the environment plays a big role on the process of star-formation suppression, some recent studies have shown that the environment starts being relevant only at $z < 0.5$ \citep{Hatfield2017}, when the galaxies have already gotten close enough to trigger the ram-pressure stripping, originating a fast quenching process to be added to an existent but slower one, dominated by strangulation \citep{Rodrigo_Munoz2019}. These two processes start to be important when the galaxy falls into the cluster, i.e. when it crosses the virial radius of a massive cluster \citep{wetzel2012}, which is a key area to understand the processes regulating the evolution of the galaxy cluster. However, to-date there are still lacking deeper studies of morphological and physical properties of galaxies in clusters, from an homogeneous multi-wavelength point of view (to mention some \citealt{liu2011} and \citealt{Yan2014}), which is the main goal of this manuscript.

In this work we aim to explore the structural (S\'ersic index, effective radius), physical (stellar masses, star-formation rates and colours, etc.) and kinematical properties of galaxies in the Hydra cluster to understand the history and evolution of its galaxies and further extend the knowledge about the effects of the environment over the galaxies. This cluster is a nearby structure located at a distance of $\sim 50$ Mpc \citep{Misgeld2011,Arnaboldi2012} and for which galaxies can be spatially well resolved, making it an ideal laboratory to fulfill our objectives. Hydra is a medium-mass compact cluster \citep{Arnaboldi2012} with a large fraction of  early-type galaxies and at least 50 ultra-compact dwarf galaxies \citep{MisgeldHilker2011,Misgeld2011}. It is classified as a type III structure \citep{BautzMorgan1970}, which means it has no dominant central member although there are two bright galaxies near to the centre: NGC 3311, a cD galaxy with radial velocity of 3825 km/s, and NGC 3309, an E3 galaxy with radial velocity of 4009 km/s (\citealt{Ventimiglia2011} and references therein). Based on X-ray data, \citet{Ventimiglia2011} have shown evidences that Hydra is a prototype of a dynamically relaxed cluster, showing an isothermal intracluster medium for the most part of the cluster region, indicating that it has not been going through any big merging process during the last few Gyrs \citep{FurushoEtAl2001,Ventimiglia2011}. Given that the cluster virial and X-ray masses are $(5.80 \pm 0.56) \times 10^{14}\,M_{\odot}$ and $(9.8 \pm 1.3) \times 10^{14}\,M_{\odot}$ respectively, and the global projected velocity dispersion is $660 \pm 52$\,km/s \citep{BabykEtAl2013} with the centre of the cluster located at $\sim$8.6 kpc northeast of NGC 3311 \citep{Barbosa2018A&A}, we will use this galaxy as the central component for practical purposes.

We perform our study from a multiwavelength point of view, by using the wide field ($\sim 2$\, sq deg) data of the Southern Photometric Local Universe Survey \cite[S-PLUS;][]{MendesdeOliveira2019} and investigate the cluster within $1 R_{200}$, the radius at which the mean density is two hundred times the critical density of the Universe. The S-PLUS provides a significant improvement in the understanding of the spectral energy distribution of galaxies at optical wavelengths due to its 12-bands filter system (5 broad and 7 narrow bands), which is a great improvement over most previous studies using up to 8 filters to perform a multi-band fitting. This is the first time that morphological analysis using 12 bands in the visible spectra is presented and the wide field of view of S-PLUS allows us to investigate in great detail any possible variation in the structural parameters and physical properties as a function of wavelength, cluster-centric distances and density.

This manuscript is presented as follows: in Section 2 we describe the data and how we construct the catalogue of Hydra Cluster galaxies, while in Section 3 we present the methodology used to estimate the structural and physical parameters. In Sections 4 and 5 we present and discuss the results to finally summarise and conclude our work in Section 6. Throughout this study we adopt a flat cosmology with $H_0=70$ km s$^{-1}$ Mpc$^{-1}$, $\Omega_M=0.3$  and $\Omega_\Lambda=0.7$ \citep{Spergel2003ApJS}.
    
\section{Data and galaxies catalogue}\label{sec:data}

Observations of the Hydra cluster were taken as part of the S-PLUS by using the T80Cam installed at the 80\,cm T80-South telescope located at Cerro Tololo Inter-American Observatory, Chile. The T80Cam has a detector with 9232$\times$9216 10$\mu$m-pixels with an effective field of view of 2 square degrees and plate-scale of 0.55 arcsec pixel$^{-1}$. S-PLUS uses the Javalambre 12-band filter system composed of 5 broad filters ($u'$, $g'$, $r'$, $i'$ and $z'$ -- $ugriz$ for simplicity) and 7 narrow-band ones (J0378, J0395, J0410, J0430, J0515, J0660 and J0861) strategically positioned in regions of the electromagnetic spectrum that have important stellar features like [O\textsc{ii}], H$\alpha$, H$\delta$, Mg$b$ and Ca triplets (see Table~\ref{tablefilter} for more information). The S-PLUS filter system was built specifically for stellar classification, yet given its spectral richness, it can also be used to analyse the physical properties of galaxies and planetary nebulae. The survey also provides photometric redshifts for galaxies brighter than 20\,mag (in $r$-band) and $z < 0.5$ \citep[see][]{Molino2019}. The observational strategy was defined to increase depth and reduce noise (the $Petrosian$ magnitude limit is $J0395 = 20.11$ -- the shallower band -- to as deep as $g = 21.79$ with $S/N \geq 3$), while dithering is used to overcome problems with bad pixels \citep{MendesdeOliveira2019}. A detailed description of the survey including filter system, calibration method (Sampedro et al. in preparation), first results and full assessment of its capabilities can be found in \citet{MendesdeOliveira2019} where the survey is described.

 \begin{table}
 \caption{S-PLUS filter system and exposure times for Hydra pointings.}
 \begin{center}
 \label{tablefilter}
 \begin{tabular}{lcccc}
 \hline
 \hline
 Filter	&	$\lambda_{\rm eff} $	&	$\Delta \lambda$	& Exp. Time &		Comment	\\	
 name	&	[\AA]	&	[\AA]	& [s]		\\	\hline
 uJAVA	&	3536	&	352	&   681 & Javalambre $u$	\\	
 J0378	&	3770	&	151	&   660 &	$[\mathrm{O}\,\textsc{ii}]$	\\	
 J0395	&	3940	&	103	&   354 & Ca H+K	\\	
 J0410	&	4094	&	201	&   177 & H$\delta$	\\	
 J0430	&	4292	&	201	&   171 & G-band	\\	
 gSDSS	&	4751	&	1545 &	99  & SDSS-like $g$	\\	
 J0515	&	5133	&	207	&   183 & Mgb Triplet	\\	
 rSDSS	&	6258	&	1465 &  120 & SDSS-like $r$	\\	
 J0660	&	6614	&	147	&   870 & H$\alpha$	\\	
 iSDSS	&	7690	&	1506 &	138 & SDSS-like $i$	\\	
 J0861	&	8611	&	408	&	240 & Ca Triplet	\\	
 zSDSS	&	8831	&	1182 &	168 & SDSS-like $z$	\\	
 \hline
 \hline
 \end{tabular}
 \end{center}
 \end{table}  

\begin{figure}
    \includegraphics[width=\columnwidth]{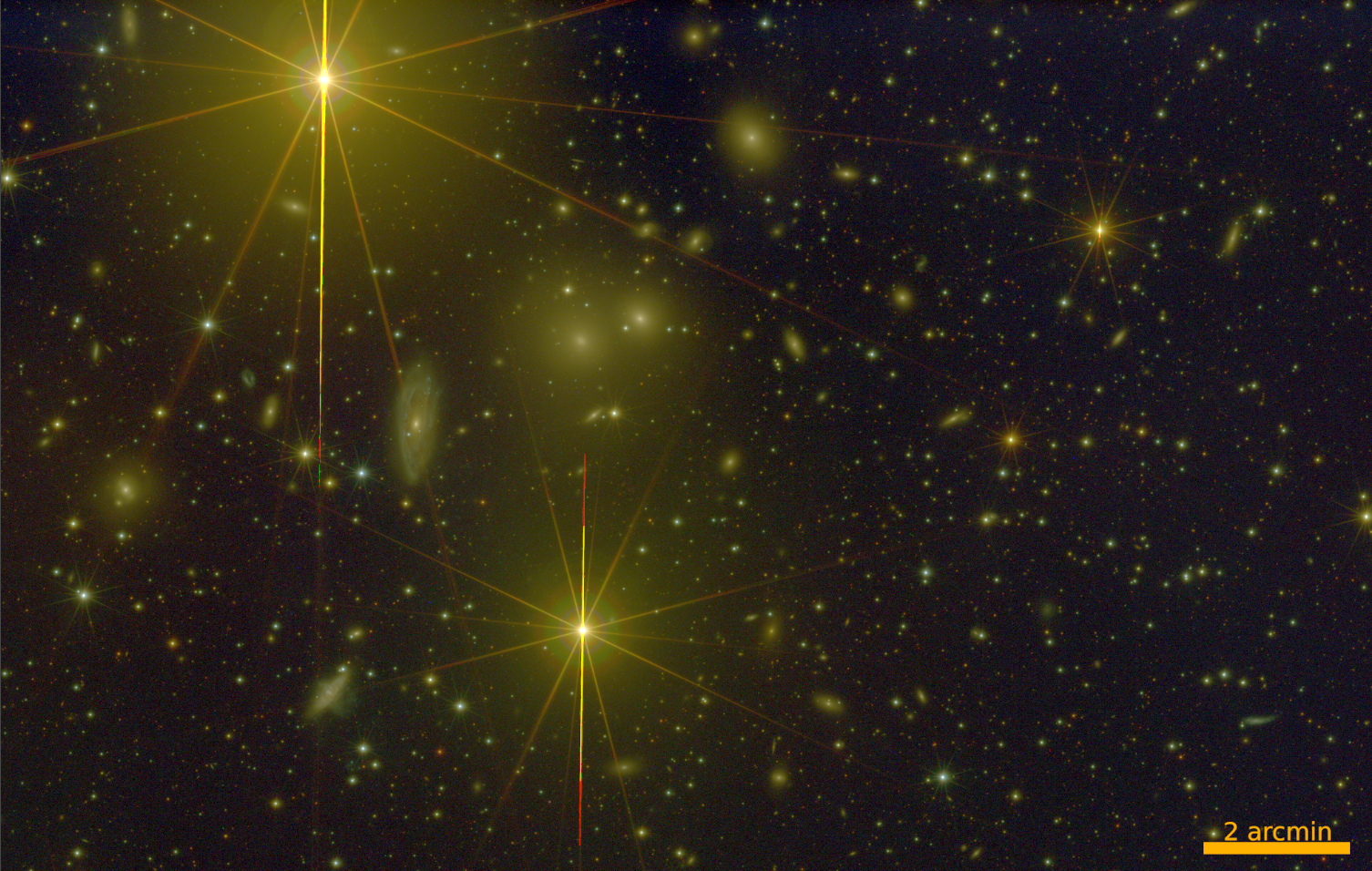}
    \caption{ Hydra central region. Colour composition using the images of the 12 S-PLUS bands with blue corresponding to $u + J0378 + J0395 + J0410 + J0430$, green to $g + J0515 + r + J0660$ and red to $J0861 + i + z$ made using the software Trilogy \citep{Coe2012ApJ...757...22C}.}
    \label{fig:composit_image}
\end{figure}

In this work we use four S-PLUS fields, covering an area of 1.4 Mpc radius, centred on the Hydra cluster, whose central region is shown in Fig.~\ref{fig:composit_image}. Table~\ref{tablefields} lists the central coordinates of each field.

%%%%%%%%%%%%%%%%%%%%%%%%%%% FIG %%%%%%%%%%%%%%%%%%%%%%%%%%%
%%%%%%%%%%%%%%%%%%%%%%%%%%% FIG %%%%%%%%%%%%%%%%%%%%%%%%%%%

%%%%%%%%%%%%%%%%%%%%%%%%%%% TAB %%%%%%%%%%%%%%%%%%%%%%%%%%%
\begin{table}
\caption{Central position J2000 of the four S-PLUS fields used in this work.}
\label{tablefields}
\begin{center}
\begin{tabular}{lllll}
\hline
\hline
FIELD    & 1      & 2      & 3      & 4      \\
\hline
RA$^{\circ}$  (J2000)  & 157.90 & 159.47 & 159.30 & 157.70 \\
DEC$^{\circ}$  (J2000) & -26.69 & -26.69  & -28.08 & -28.08 \\
\hline
\hline
\end{tabular}
\end{center}
\end{table}
%%%%%%%%%%%%%%%%%%%%%%%%%%% TAB %%%%%%%%%%%%%%%%%%%%%%%%%%%

\subsection{Kinematic selection of Hydra cluster galaxies}\label{subsec:members}

We select all galaxies that are gravitationally bound to the cluster and inside $1\,R_{200}$. First we select all galaxies with peculiar velocities lower than the cluster escape velocity, which is calculated in the line-of-sight relative to the cluster recessional velocity, as defined by Eq.~\ref{eq:pec_vel} \citep[see][]{Harrison1974ApJ,Jaffe2015}: 
\begin{equation}
  pec = c\frac{z-z_{cl}}{1+z_{cl}}
	\label{eq:pec_vel}
\end{equation}

where $z_{cl} =  0.012$ \citep{BabykEtAl2013} is the cluster redshift, $z$  is the redshift of each galaxy obtained from the NASA/IPAC Extragalactic Database (NED\footnote{https://ned.ipac.caltech.edu/}), and $c$ is the speed of light. The cluster escape velocity ($v_{esc}$), in km/s, is calculated using the Eq.~\ref{eq:escape_velocity} \citep[Eq.1 in ][]{Diaferio1999}:

\begin{equation}
   v_{esc} 	\simeq 927\left(\frac{M_{200}}{10^{14}h^{-1}M_{\odot}}\right)^{1/2}\left(\frac{R_{200}}{h^{-1}\rm Mpc}\right)^{-1/2}
	\label{eq:escape_velocity}
\end{equation}

The determination of the cluster escape velocity depends on the $M_{200}$ (mass within the $R_{200}$), $R_{200}$, which in turn are determined by the velocity dispersion ($\sigma$), and $h$= $H_0$/100 km s$^{-1}$ Mpc$^{-1}$. To obtain $\sigma$ we use only galaxies with radial velocities ranging from 1800 to 6000\,km/s, suggested by \citet{Ventimiglia2011}. We found a biweight velocity dispersion of $690 \pm 28$\,km/s as defined by the Eq.~\ref{eq:velocity_disperion}:

\begin{equation}
   \sigma_{BI}^{2} = N\frac{\sum_{_{\left|u_{i} \right |<1}} (1-u_{i}^{2})^{4}(v_{i}-\overline{v})^{2}}{D(D-1)}
	\label{eq:velocity_disperion}                        
\end{equation}

where $v_{i}$ is the peculiar velocity and $\overline{v}$ is its average, as described in \citet{Beers1990AJ} and \citet{Ruel2014}, and N is the number of members. $D$ is given by Eq.~\ref{eq:D_velocity_disperion}.

\begin{equation}
   D =  \sum_{\left|u_{i} \right |<1}(1-u_{i}^{2})(1-5u_{i}^{2})
	\label{eq:D_velocity_disperion}
\end{equation}
where $u_{i}$ is defined as shown by Eq.~\ref{eq:u_velocity_disperion}.
 
\begin{equation}
   u_{i} =  \frac{v_{i}-\overline{v}}{9\rm{MAD}(v_{i})}
	\label{eq:u_velocity_disperion}
\end{equation}
and $\rm{MAD}(v_{i})$ is the median absolute iation of the velocities. The biweight velocity dispersion uncertainty is calculated as given by Eq.~\ref{eq:err_velocity_disperion}.

\begin{equation}
   \Delta\sigma_{BI} =\frac{C_{BI}\sigma_{BI}}{\sqrt{N-1}}
	\label{eq:err_velocity_disperion}
\end{equation}
where $C_{BI} = 0.92$. After measuring $\sigma$ ($690 \pm 28$\,km/s in our case), we can use it to determine  $M_{200}$ and $R_{200}$ following \citet{LeonardKing2010}. They use a singular isothermal sphere model profile, assume spherical symmetry, and that the 3D velocity dispersion can be described from the line-of-sight one dimensional velocity dispersion \citep{Gonzalez2018}. The relationship found by \citet{LeonardKing2010} between $\sigma$, $M_{200}$ and $R_{200}$ is as follows:

\begin{equation}
  M_{200} = \frac{2\sigma^{3}}{\sqrt{50}GH}
	\label{eq:m200}
\end{equation}

\begin{equation}
  R_{200} = \frac{\sigma}{\sqrt{50}H}
	\label{eq:r200}
\end{equation}
where $G$ and $H$ are the gravitational and Hubble constants, respectively. We find a $M_{200} = 3.1 \pm 0.4\times10^{14}\,M_{\odot}$ and $R_{200} = 1.4 \pm 0.1$\,Mpc. We also calculate $M_{200}$ using the relation between velocity dispersion and $M_{200}$, as obtained by \citet{Munari2013} for simulated galaxy clusters, and described by Eq.~\ref{eq:M200_Munari}. 

\begin{equation}
  \frac{\sigma_{D1}}{\text{km s}^{-1}} = A_{1D} \left [ \frac{h(z)M_{200}}{10^{15}M_{\odot}}\right ]^{\alpha}
	\label{eq:M200_Munari}
\end{equation}

The values of $A_{1D}$ and $\alpha$ for the Eq.~\ref{eq:M200_Munari} are found in \citet{Munari2013}. The velocity dispersion in the simulation is determined using Dark Matter particles (DM), subhaloes (SUB) and galaxies (GAL). The three cases consider the contribution of AGN feedback. Using the Eq.~\ref{eq:M200_Munari} we obtained the following $M_{200}$ masses, in units of $10^{14}\,M_{\odot}$, for DM, SUB and GAL, respectively, $3.5 \pm 0.4$, $3.1 \pm 0.3$ and $3.3 \pm 0.3$, which are in agreement with $M_{200}$ estimated using Eq.~\ref{eq:m200}, as well as other reported values \citep[e.g. $M_{200} = \,3.8 \times 10^{14}$ from][]{Comerford2007}.

Using the calculated $M_{200} = 3.1 \pm 0.4\times10^{14}\,M_{\odot}$ and $R_{200} = 1.4 \pm 0.1$\,Mpc, we obtain $v_{esc} = 1379$\,km/s from Eq.~\ref{eq:escape_velocity}, which gives us a sample of 193 galaxies satisfying the criteria to be members of the Hydra Cluster within $1\,R_{200}$. To check our consistency, we use another method to select Hydra members by applying a 3 sigma-clipping to the recessional velocity distribution of the galaxies. For this, we reject galaxies with recessional velocities larger than $3\,\sigma$ which results in a sample of 223 objects located within $1\,R_{200}$.

 Although both methods provide  numbers of cluster members within the same order of magnitude, ensuring that our choice will not affect the results of this paper, we will use the sample selected by the first method to ease further comparisons with other studies.
 
\subsection{Final complete sample of Hydra galaxies}

In order to produce a complete galaxy sample in Hydra, we performed a cross match between different spectroscopic surveys in the Hydra area from \citet{Richter1987A&AS,Stein1996A&AS,Jones2004MNRAS}. All these spectroscopic studies are complete for galaxies brighter than $\sim$ 16 mag in the r-band. Therefore, in this study  we only include the selected galaxies in sect.~\ref{subsec:members} that are brighter than 16 mag in the r-band, to ensure a complete sample of galaxies analysed. Given this, all galaxies analysed here have large S/N, between 24 and 145 (in the r-band) with a mean value of 55.

\begin{figure}

\includegraphics[width=\linewidth]{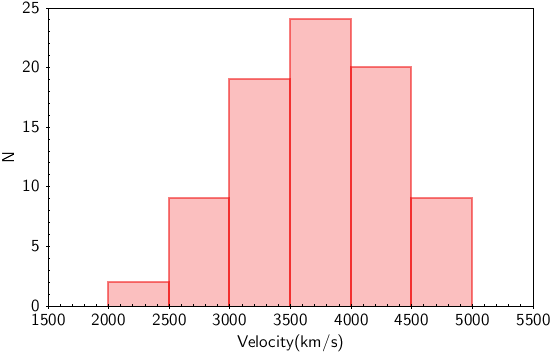}
    \caption{Histogram of the velocity distribution for the 81 Hydra Cluster galaxies of our sample. The y-axis shows the number of galaxies at each velocity bin.}
    \label{fig:hits_belocity_H}
\end{figure}

\begin{figure} % figura perdida no lugar do paper
    \includegraphics[width=\columnwidth]{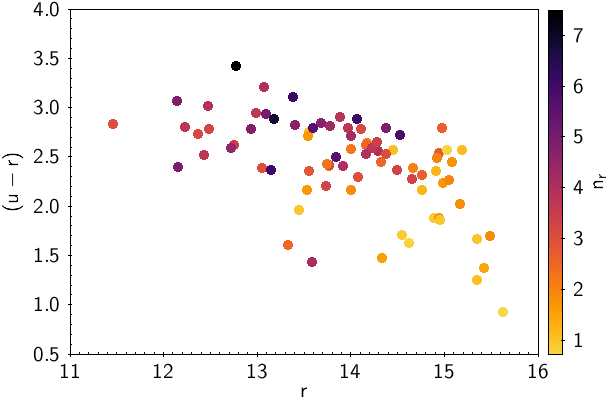}
    \caption{Colour magnitude diagram for the galaxies in Hydra Cluster. The dots are colour-coded by the S\'ersic index in the r-band. Most of the galaxies are located within the red sequence. There is a clear relation between the S\'ersic index and the galaxy's colour.}
    \label{fig:CMD}
\end{figure}

We note that some of the galaxies of our sample were not well fitted by MegaMorph (see Section \ref{sec:morphology}) due to either foreground stars contamination or to its proximity to the CCD edge. After removing these objects we obtain a final sample of 81 galaxies brighter than 16 ($r$-band), within $1\,R_{200}$, and with peculiar velocities lower than $v_{esc} = 1379$ km/s. We use these galaxies to analyse the behaviour of the morphological parameters, for example S\'ersic index. Figures~\ref{fig:hits_belocity_H} and \ref{fig:CMD} show the distribution of recessional velocities and a colour-magnitude diagram (CMD)  respectively, for the selected 81 galaxy members. We note, in Fig.~\ref{fig:CMD}, that most galaxies are located in the red sequence, and there is a clear relation between the S\'ersic index and the galaxy's colour, redder galaxies have a higher S\'ersic index.

%%%%%%%%%%%%%%%%%%%%%%%%%%% FIG %%%%%%%%%%%%%%%%%%%%%%%%%%%
%%%%%%%%%%%%%%%%%%%%%%%%%%% FIG %%%%%%%%%%%%%%%%%%%%%%%%%%%

\section{Methodology}
\subsection{Morphological Parameters}\label{sec:morphology}

The morphological classification of galaxies can be done by visual inspection \citep[e.g.,][]{Lintott2008,Kartaltepe2015,Simmons2017}. It is, however, extremely costly on human resources to perform over millions of galaxies which are available within modern surveys. For this volume we must rely on the computational power we have available and use a different approach to obtain the structural parameters of galaxies, such as S\'ersic index ($n$), effective radius ($R_{e}$), bulge-to-total Flux ($B/T$), Gini coefficient and the second-order moment of the brightest 20 percent of the galaxy \citep{Lotz2004AJ....128..163L}. All these parameters can be used to classify galaxies morphologically, specially considering that morphological analyses can be done from the infrared to the ultraviolet regions of spectra \citep{GildePaz2007,Lotz2008,Wright2010,Dobrycheva2017}. Indeed, the S\'ersic profile describes how the intensity of a galaxy varies with radius, providing information regarding the morphology of the galaxy \citep{Sersic1963}. This is the approach we adopt in this work, whose motivation is two folded: on one side, automatic classification allows us to consider a  morphological analyses from the infrared to the ultraviolet regions of spectra, done in a consistent way. On the other hand, it will allow us to readily compare our results with future S-PLUS and J-PLUS data, which will deliver multiband data for millions of galaxies \citep{MendesdeOliveira2019},  as well as with other work, that have also performed automatic classification.

With the purpose of estimating the S\'ersic index ($n$), effective radius ($R_{e}$) and the magnitudes ($m$) for the members of the Hydra Cluster in all S-PLUS bands, we used the code \textsc{MegaMorph-GALAPAGOS2} \citep{Bamford2011,Vika2013,Haubler2013}. This code performs  a multiwavelength two-dimensional fitting using the algorithm \textsc{GALFITM} \citep{Peng2002,Peng2010,Vika2013}. \textsc{GALFITM} extracts structural components from galaxy images by modelling the surface-brightness with different profiles: Nuker law \citep{Lauer1995}, the S\'ersic profile, exponential disc, Gaussian or Moffat functions \citep{Moffat1969A&A}. The main advantage of a simultaneous multiwavelength fitting is an increasing in the accuracy of the estimated parameters \citep{Vika2015}. In addition, \textsc{MegaMorph} allows us to fit all galaxies in a given field simultaneously, which is of great convenience. We note that \textsc{MegaMorph} was used to perform a multiband fitting in several previous studies \citep{Vika2013,Vika2014MNRAS,Vika2015,Haubler2013,Vulcani2014,Kennedy2015,Dimauro2018MNRAS}. In addition, it was tested with simulated galaxies \citep{Haubler2013}. Thus it is a well-tested code. The galaxies analysed here have large S/N, between 24 and 145 (in the r-band) with a mean value of 55, and good imaging quality, since all images were taken under photometric conditions; thus we trust that the output parameters provided by MegaMorph are reliable. Nevertheless, as a sanity check, we tested how well \textsc{GALFITM} recovers the galaxy parameters for our particular data set of S-PLUS. We present this test in the Appendix~\ref{appendix:simulation}, where we generate simulated galaxies with the same features, S/N, filters and background levels as those in the S-PLUS fields of Hydra, and find that \textsc{GALFITM} can retrieve the input galaxy's parameters with high reliability, within a percentage error of $\sim4$ percent (see Appendix~\ref{appendix:simulation} for more details).

In this work, the centre of each galaxy was determined by SourceExtractor using a deep detection image, generated for each field as a weighted combination of the $g$, $r$, $i$ and $z$ broad-band images. Then, for each galaxy, we fit a single S\'ersic profile simultaneously for images in all filters, by fixing for each filter the central position measured on the detection image. Each parameter to be measured is modeled as a function of wavelength and the degrees of freedom afforded to model each of these parameters is determined by a set of
Chebyshev polynomials. This analysis provided us structural parameters, such as the S\'ersic index and effective radius (modelled as quadratic functions of wavelength) as well as b/a and position angle (modelled as linear functions of wavelength) \citet{Haubler2013}. The S\'ersic profile, which is modeled for each galaxy, is define in Eq.~\ref{eq:Sersic_profile}.

\begin{equation}
   I(r)=I_{e}\left \{ -b_{n}\left [ \left (\frac{r}{R_{e}}  \right )^{\frac{1}{n}}-1 \right ] \right \}
	\label{eq:Sersic_profile}
\end{equation}

where $I_{r}$ is the intensity, $R_{e}$ and $I_{e}$ are the effective radius and the intensity inside an $R_{e}$, respectively. $n$ is the S\'ersic index, also known as the concentration parameter, and $b_{n}$ is a function of $n$ that satisfies $\Gamma(2n)=2\gamma(2n,b_{n})$. In cases where the galaxy profile is more concentrated, $n$ presents higher values. For instance, with $n$ = 4 we obtain the well known de Vaucouleurs profile and it corresponds to the typical profile of an elliptical galaxy. For $n$=1, we have a typical profile of an exponential disk, and if $n<1$ this is probably due to the presence of a bar \citep{Peng2010}.

In order to determine the best fit, \textsc{GALFITM} uses a Levenberg-Marquardt technique, which finds the optimum model by minimising the $chi^{2}$. In Fig.~\ref{fig:G004} we present the \textsc{GALFITM} output for the galaxy ESO 437- G 004, as an example, where observed images, models and residuals (observed minus models) are shown in top, middle and bottom panels, respectively. We can easily see the spiral arms and substructures in the residual images of this example.

Magnitudes and Sersic index for the 81 analysed galaxies are available in Tables \ref{tab:Mag_MegaMorph} and \ref{tab:Sersic_MegaMorph} (see Appendix~\ref{appendix:simulation} for a complete version of these tables). All magnitudes used in this work have been corrected for Galactic extinction, by using a \citet{CCM1989} law and the maps from \citet{Schlegel1998ApJ}.

\begin{figure*}
\label{fig:G004}
\includegraphics[width=\textwidth]{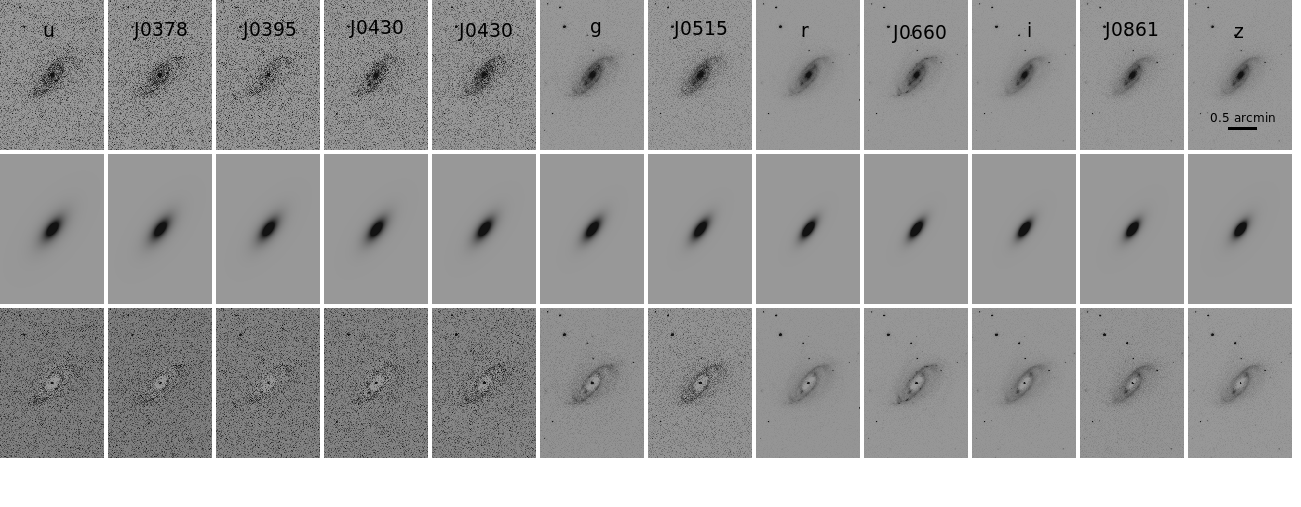}
    \caption{Galaxy ESO 437-G 004 as observed by S-PLUS (top panels), the models calculated using \textsc{GALFITM} (middle panels), and the residual image (observed minus modelled -- bottom panels).}
    \label{fig:G004}
\end{figure*}

\subsection{ Estimation of stellar masses}\label{sec:mass}

\subsubsection{Stellar mass from colours}\label{sec:masscolours}

The stellar mass is one of the main properties of a galaxy and it is well correlated with its luminosity \citep{Faber1979ARA&A}. Different types of galaxies increase their stellar masses at different rates, starburst galaxies form stars at higher rates than main sequence galaxies \citep{Papovich2005ApJ}. \citet{Bell2003}, using a large sample of 22,679 galaxies observed with the Two Micron All Sky Survey \citep{Skrutskie1997ASSL} and the SDSS, calculated stellar mass-to-light ratio $M_{\star}/L$ as a function of the colour $(g - i)$ assuming a Salpeter initial mass function \citep[IMF,][]{Salpeter1955}. In this work we use the $i$-band luminosities, along with the colour $(g - i)$, to estimate the stellar mass following the definitions given by \citet{Bell2003}, which are shown in the Eqs.~\ref{eq:Bell_g_i}. 

\begin{equation}
   \log\ M_{\star}/L_{i} = -0.152 +0.518\times(g-i)
	\label{eq:Bell_g_i}
\end{equation}

We found a median uncertainty of 0.06 dex for the estimated stellar masses of our Hydra galaxies. Using a sample of galaxies from the Galaxy And Mass Assembly (GAMA), \citet{Taylor2011} have shown that the colour $(g-i)$ can be used to infer $M_{\star}/L_{i}$ with a typical uncertainty of < 0.1~dex following the definition shown in Eq.~\ref{eq:Taylor_g_i}

\begin{equation}
   \log\ M_{\star}/L_{i} = 1.15 + 0.7\times(g-i) -0.4\times M_{i}
	\label{eq:Taylor_g_i}
\end{equation}

where $M_{i}$ is the absolute magnitude in the $i$-band. The authors used the SSPs models from \citet{BruzualCharlot2003} and assumed a \citet{Chabrier2003} IMF with a \citet{Calzetti2000} extinction law. In this work we also use the relation presented in \citet{Taylor2011} to estimate the stellar masses.

\subsubsection{Stellar mass using LePHARE}

Another way to estimate the stellar mass of a galaxy is using a Spectral Energy Distribution (SED) or spectral fitting code \citep{Bruzual1993,Cid2005MNRAS,daCunha2012}. These codes use models to fit the SED, allowing us to determine the ages and metallicities of the stellar populations. Then, it is possible to obtain several physical parameters of the galaxy, including its stellar mass.

Following this approach, we perform a SED fitting for the galaxies detected inside of the S-PLUS field containing the centre of the Hydra Cluster (only one field for a consistency check). The SED fitting process was done using the code PHotometric Analysis for Redshift Estimate \citep[\textsc{LePHARE}][]{Arnouts1999,Ilbert_et_al_2006} and the stellar population libraries of \citet{BruzualCharlot2003} with a \citet{Chabrier2003} IMF. The models have 3 metallicities $0.2\,Z_\odot$, $0.4\,Z_\odot$ and $1\,Z_\odot$ with age ranging from 0.01 to 13.5\,Gyrs. 

\begin{figure}
\includegraphics[width=\linewidth]{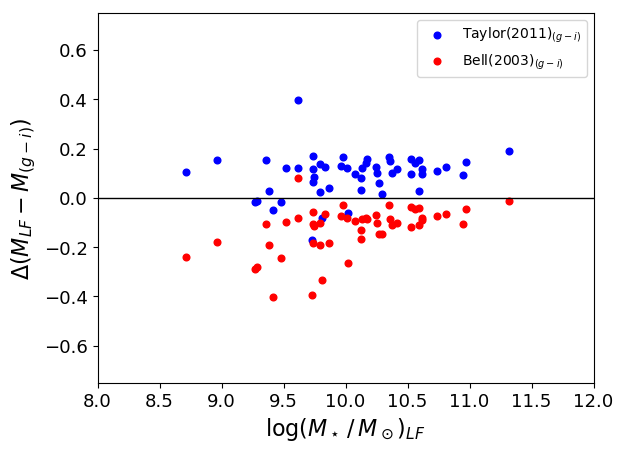}
    \caption{Comparison between the stellar masses estimated using colours and using the code \textsc{LePHARE}. The y-axis shows the difference between the stellar masses calculated with \textsc{LePHARE} and colour relation. In red and blue are the masses estimated using the colour $(g-i)$ from \citet{Bell2003} and \citet{Taylor2011}, respectively.}
    \label{fig:massBT}
\end{figure}

In Fig.~\ref{fig:massBT} we compare the resulting masses obtained by the SED fitting method with respect to the masses derived from luminosities and colours (as described in Section~\ref{sec:masscolours}). Red dots are the stellar masses estimated using the colour relation $(g-i)$from \citet{Bell2003}. We have scaled the Bell mass-luminosity relation by 0.093 dex to take into account the use of Salpeter IMF rather than a Chabrier IMF, as described in \citet{Taylor2011}. Blue dots are the stellar masses estimated using the colour relation $(g-i)$ from \citet{Taylor2011}. The stellar masses estimated using \citet{Bell2003} are on average 0.22~dex larger than those estimated via \citet{Taylor2011}. When the stellar masses derived from the SED fitting are compared with those derived from the colour relations we find, on average, a percentage error of 1.2 and 2.0 for the masses estimated using the \citet{Taylor2011} and \citet{Bell2003} relations, respectively. In both cases a linear regression of the stellar masses obtained from the two methods provides a good fit, with a coefficient of determination $R^{2} = 0.97$, indicating a good lineal correlation between the data.
The coefficient of determination is defined as  $R^{2}$ = 1 -  ($SS_{res}$/$SS_{tot}$), where $SS_{res}$ is the sum of residual errors and $SS_{tot}$ is the total errors. Its value ranges from 0 to 1, where 1 corresponds to the maximum correlation. 

Given that the two colour relations show a good correlation with the masses estimated using a SED fitting, and considering that the masses estimated using colours is less time consuming, we therefore use the stellar masses for all the S-PLUS fields obtained from the colour relation by \citet{Taylor2011}, which shows the smallest percentage error.

\begin{figure}
\includegraphics[width=\linewidth]{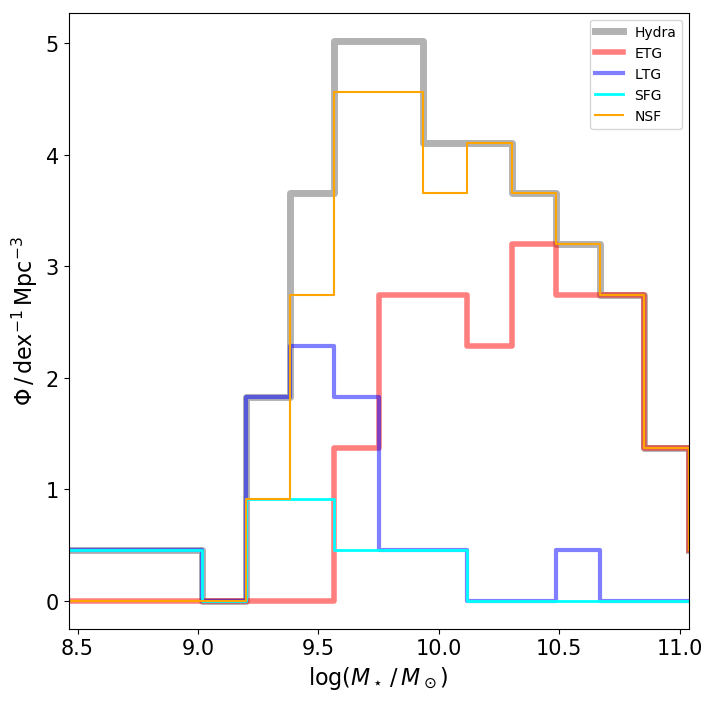}
    \caption{Stellar mass function of galaxies brighter than m$_r$=16 within a sphere with 1.4Mpc of radius centred on Hydra. The grey line is the 81 galaxies belonging to Hydra  selected as indicated in subsection~\ref{subsec:members}. The orange and cyan lines are the quenched (NSFGs) and SFGs. The red and blue lines are ETGs  and LTGs respectively.}
    \label{fig:SMF}
\end{figure}

We show the stellar mass function (SMF) of the Hydra galaxies analysed in this work as a grey histogram in Fig.~\ref{fig:SMF}. The SMF is dominated by ETGs and quenched galaxies, as defined in the following sections. Our sample is complete for stellar masses  $ \geq  3.3 \times10^{9}M_{\odot}$, which was determined from the colour relation using the faintest magnitude (16 mag in the r-band) and the reddest colour of our sample of galaxies.  We can see from Fig.~\ref{fig:SMF} that there are virtually no ETGs and quenched galaxies (NSFG) in the lower mass end of the SMF. The stellar masses are listed in Table~\ref{tab:parameter} of the Appendix.

\subsection{The star formation rate estimation}\label{sec:sfr}

To further extend our understanding of the galaxies inside the Hydra Cluster and the influence of the environment over the evolutionary path of the members, we estimate the star formation rate ($SFR$) and the specific star formation rate ($sSFR$) for the galaxies of our sample. For this task we use the $J0660$ narrow band filter, which is centred around the rest frame wavelength of H$\alpha$. We note that there are several advantages in using photometric data to determine H$\alpha$ emission in galaxies. First, the use of S-PLUS data allow us to perform an homogeneous analysis on the H$\alpha$ emission of galaxies in Hydra, where all the data were observed under similar conditions. In addition, imaging surveys are not biased by the orientation of the slits, as in the case of spectroscopic studies. 
Furthermore, the S-PLUS will cover a huge area in the sky, and in the near future the photometric redshifts will be available, increasing the number of objects belonging to Hydra for which we will have S-PLUS data. In this way, the same criteria will be used to select and analyse galaxies providing homogeneous data with S-PLUS.

We select the emission line galaxy candidates in our sample based on two criteria: one regarding the equivalent width and the other one regarding a given colour excess for the objects  (known as 3$\Sigma$ cut). For the galaxies to be considered as line emitters, the H$_\alpha$ equivalent width (EW$_{J0660}$) should be greater than 12 \AA. This is following \citet{Vilella2015}, who determine that J-PLUS cannot resolve, with a precision of 3$\sigma$, EW$_{J0660}$  < 12 \AA. In this work, we use the same criteria given that S-PLUS and J-PLUS use twins telescopes, with the same filters systems, thus we can utilize the same criteria to analyse the galaxies if the objects have the similar photometric conditions. We use the equation \ref{eq:equivalent_with} to determine the galaxies’s EW$_{J0660}$.

\begin{equation}
   EW_{J0660} = \Delta_{J0660} (Q-1)\frac{Q-1}{1-Q\epsilon}
	\label{eq:equivalent_with}
\end{equation}

where $\epsilon \equiv \Delta_{J0660}/\Delta_{r}$  and  $m_{r}$ - $m_{J0660} = 2.5\log Q$. The $m_{r}$ and $m_{J0660}$ are the apparent magnitudes.

\begin{figure}
\includegraphics[width=8.0cm,height=6cm,keepaspectratio]{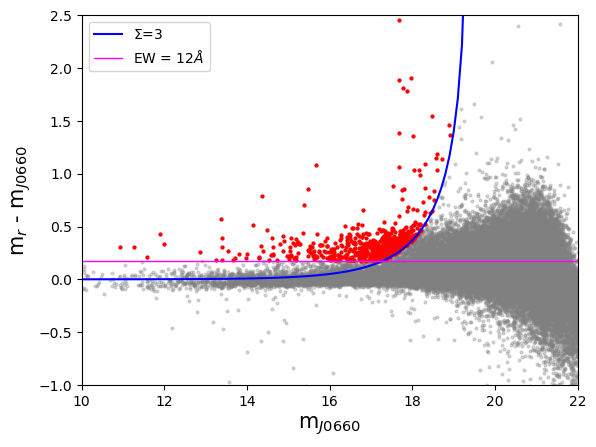}
    \caption{$m_{J0660}$ excess as a function of $m_{J0660}$ magnitude. The blue line represent the $\Sigma$ cut of 3. The magenta horizontal line present the EW$_{J0660}$ cut of 12 \AA. The grey points represent all $m_{r}-m_{J0660}$ detected sources. The red points are the sources that can be considered to have a narrow-band excess in the field.}
    \label{fig:colour_excess}
\end{figure}

Emission line galaxies are objects with colour excess greater than zero (r - J0600)  > 0. However, to quantify the colour excess when compared to a random scatter expected for a source with zero colour, we use a 3$\Sigma$ cut \citep{Sobral2012MNRAS}. We use Eq.~\ref{eq:sigma_cut} to define the 3$\Sigma$ curve \citep{Khostovan2020MNRAS} for our second selection criterion

\begin{equation}
   \Sigma =1 - \frac{10^{-0.4(m_{r} - {m_{J0660}})}}{10^{ZP - m_{J0660}}\sqrt{\sigma_{J0660}^{2} +\sigma _{r}^{2}} }
	\label{eq:sigma_cut}
\end{equation}

where $\sigma_{J0660}$ and $\sigma_{r}$ are the rms error. SourceExtractor \citep{Bertin1996} was used in this case to obtain the galaxy magnitudes, and their errors, for each Hydra's field, with a AUTO aperture around the source.  ZP is the photometric zero-point for J0660 filter. Figure \ref{fig:colour_excess} shows the application of these selection criteria to the detected sources in one Hydra field. Only galaxies that obey the 3$\Sigma$ cut and EW$_{J0660}$ > 12 {\AA}  were selected as line emitters. We apply the same selection on all 4 S-PLUS fields analysed in this work and find only 10 galaxies in Hydra that meet our selection criteria to be emission line galaxy candidates. These galaxies have $H_{\alpha}$ luminosities between $1.4\times10^{40}$ and $1.5\times10^{41}$ erg/s.

To measure the H$\alpha$ flux on the emission-line selected galaxies, we apply the Three Filter Method \citep[TFM][]{Pascual2007} -- which has already been successfully implemented to similar data from the J-PLUS survey \citep{Cenarro2019,Vilella2015} -- using the $r$ and $i$-bands to trace the linear continuum and $J0660$ to contrast the line. We obtain a ``pure'' H$\alpha$ plus [NII] doublet emission. Even though it is impossible to separate the contribution of the [NII] ~lines from the H$\alpha$ flux, given the width of the $J0660$ filter, we can use the empirical relations from \citet{Vilella2015} to estimate the level of contamination one could expect in such kind of data. Following the relations from the Eq.~\ref{eq:NII_cor}, we are able to subtract the contribution of [NII] for the flux and obtain a pure H$\alpha$ emission for our galaxies.

\begin{equation}
    \log F_{H\alpha}= 
    \begin{cases}
    0.989\log F_{H\alpha+[NII]} - 0.193, & \text{if $g-i \leq 0.5$} \\
    0.954\log F_{H\alpha+[NII]} - 0.753, & \text{if $g-i > 0.5$}
	\end{cases}
	\label{eq:NII_cor}
\end{equation}

The analysis described above allow us to determine a sensitivity of the $F_{H\alpha}$ is 1$\times10^{-13}\ \mathrm{erg\,s^{-1}cm^{-2}}$ which corresponds to a surface brightness of 21\,mag/arcsec$^{-2}$.
We use the H$_{\alpha}$ flux to determine the H$_{\alpha}$ luminosity, and then the classical relation proposed by \citet{Kennicutt1998} was used to estimate the SFR. The SFRs obtained from this relation must be corrected for dust attenuation. A common correction arises from the assumption of an extinction A(H$\alpha$) = 1 mag, as proposed by \citet{Kennicutt1992ApJ}. However, this correction overestimates the SFR for galaxies with low H$_{\alpha}$ luminosities \citep[H$_{\alpha}$ luminosity of < $10^{40.5}$ ergs s$^{-1}$,][]{Ly2012ApJ}, which is the case for some galaxies analyzed in this work. For these reasons, we choose to use the relation between the intrinsic and observed SFR (corrected by the obscuration), which is presented in \citet{Hopkins2001AJ} and updated by \citet{Ly2007ApJ}. The relation is shown in Eq.~\ref{eq:SFR_corre}.

\begin{equation}
\begin{split}
   \log SFR_{obs}(H_{\alpha})=\log SFR_{int} - 2.360 \\
   \times \log\left [ \frac{0.797\log(SFR_{int})+3.786}{2.86} \right ]
	\label{eq:SFR_corre}
\end{split}
\end{equation}

where SFR$_{int}$ and SFR$_{obs}$ are the intrinsic and observed SFR respectively. 

Finally, $sSFR$ were estimated by using SFR and the stellar masses derived in section ~\ref{sec:masscolours}. Table \ref{tab:parameter} in the appendix lists the $sSFR$ (column 6). A $sSFR$ threshold is defined empirically to separate star-forming galaxies (SFGs) from the non-star forming ones \citep[NSFGs][] {Weinmann2010}. We consider a galaxy as star-forming if its $sSFR > 10^{-11}\,\mathrm{yr}^{-1}$ otherwise it will be classified as quiescent following the threshold commonly used in the literature \citep[see for example][]{wetzel2012,Wetzel2013,Wetzel2014}. We note, however, that some other studies use different thresholds for the $sSFR$ at different redshifts to separate SFG from NSFG, as for example \citealt{Lagana2018} and \citealt{Koyama2013}. Based on these definitions we found that the Hydra cluster has 88\,percent of the galaxies already quenched.

\section{Results}

In this section we present the morphological classification of Hydra galaxies as early and late-type galaxies using $n$ and colours. We then analyse the spatial distribution of the different types of galaxies in terms of their physical and structural parameters as well as their behaviour with respect to the cluster central distance and with the cluster density. We also present the $n$ behaviour as a function of the 12 S-PLUS filters. In addition, we analyse the phase-space diagram and we explore the presence of substructures in Hydra cluster

\subsection{Early-type and late-type galaxies classification}\label{subsec:ETG_LTG}

Based on the physical properties of the stellar populations of ETGs and LTGs, it is straightforward to separate between these two populations using a colour-magnitude diagram \citep{Bell2004ApJ}. ETGs have in general much older and redder stellar populations, which allocates this population in a very well determined position of the diagram separated from the bluer and star-forming LTGs \citep{Lee2007}. Using \textit{SDSS} data \citet{Vika2015} combined $n_{r}$ and the colour cut $(u - r) = 2.3$ to separate these two galaxy classes (see also \citealt{Park2005} for different values of the colour cut). \citet{Vika2015} classified the galaxies with $(u - r) \geq 2.3$ and $n_{r} \geq 2.5$ as ETGs and galaxies with  $(u - r) < 2.3$ and $n_{r} < 2.5$  as LTGs. Following these parameters to classify our sample, $\sim54$ percent of the galaxies (44 objects) are ETGs, whereas $\sim 23$ percent (19 galaxies) are LTGs.

\begin{figure}
\includegraphics[width=8.0cm,height=6cm,keepaspectratio]{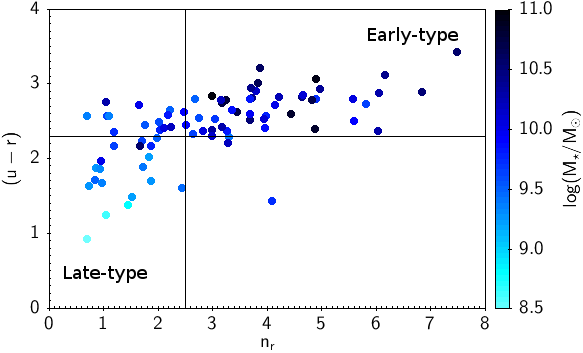}
    \caption{Results of the morphological fitting, performed with MegaMorph-GALAPAGOS2. The x-axis shows the S\'ersic index in the $r$-band the y-axis displays the galaxies' $(u - r)$ colour. Colours represent the stellar mass of each galaxy, as indicated by the colour bar. The vertical and horizontal lines are in $n_{r} = 2.5$ and $(u-r)$ = 2.3, respectively. }
    \label{fig:early-late}
\end{figure}

\begin{figure}
\includegraphics[width=8.0cm,height=6cm,keepaspectratio]{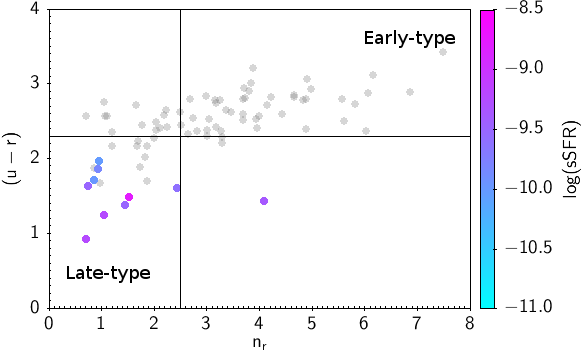}\caption{Same as Fig.~\ref{fig:early-late}, but colour-coded by the $log(sSFR)$. The grey dots are NSFG. }
    \label{fig:early-late-ssfr}
\end{figure}

In Figs. ~\ref{fig:early-late} and ~\ref{fig:early-late-ssfr} we present the $n_{r}$ versus the ($u$-$r$) colour for the whole sample used in this study, where symbols in the figures are colour-coded by stellar mass and sSFR, respectively. As expected, ETGs (top-right region on each plot) are more massive than LTGs (bottom-left region). In Fig.~\ref{fig:early-late-ssfr}, we show that LTGs are forming stars at a level of -10.0$<$log(sSFR)$<$-8.8 whereas all the ETGs are quenched. It is interesting to note that there is a population of blue + green galaxies ($(u-r)$ < 2.3) in Hydra cluster that do not present $H\alpha$ emission at the detection level. The SMF of SFGs and NSFGs is shown in the Fig.~\ref{fig:SMF}, as cyan and orange histograms respectively. We find that the SFGs are less massive than the NSFGs, where the NSFG have basically the same behaviour of the global SMF, since only 10 galaxies of the sample are SFGs. There is a mix of populations in the top left in the Figures~\ref{fig:early-late} and ~\ref{fig:early-late-ssfr}, which represents a $\sim17$ percent of our sample.

\subsection{Spatial distribution: structural and physical parameters}

In this section we analyse how the morphological and physical parameters change with respect to the distance to the cluster centre as well as a function of the projected local density. For each galaxy we estimate the projected local density defined as $\Sigma_{10} =10/A_{10}$, where  $A_{10} = \pi R_{10}^{2}$(Mpc) is the area of the circle that contains the nearest 10 galaxies and $R_{10}^{2}$ is the radio of the circle, as described by \citet{Fasano2015}. Each galaxy is in the centre of the circle.

\begin{figure}
\centering
\textbf{ETGs and LTGs}\par\medskip
\includegraphics[width=8.0cm,height=10cm,keepaspectratio]{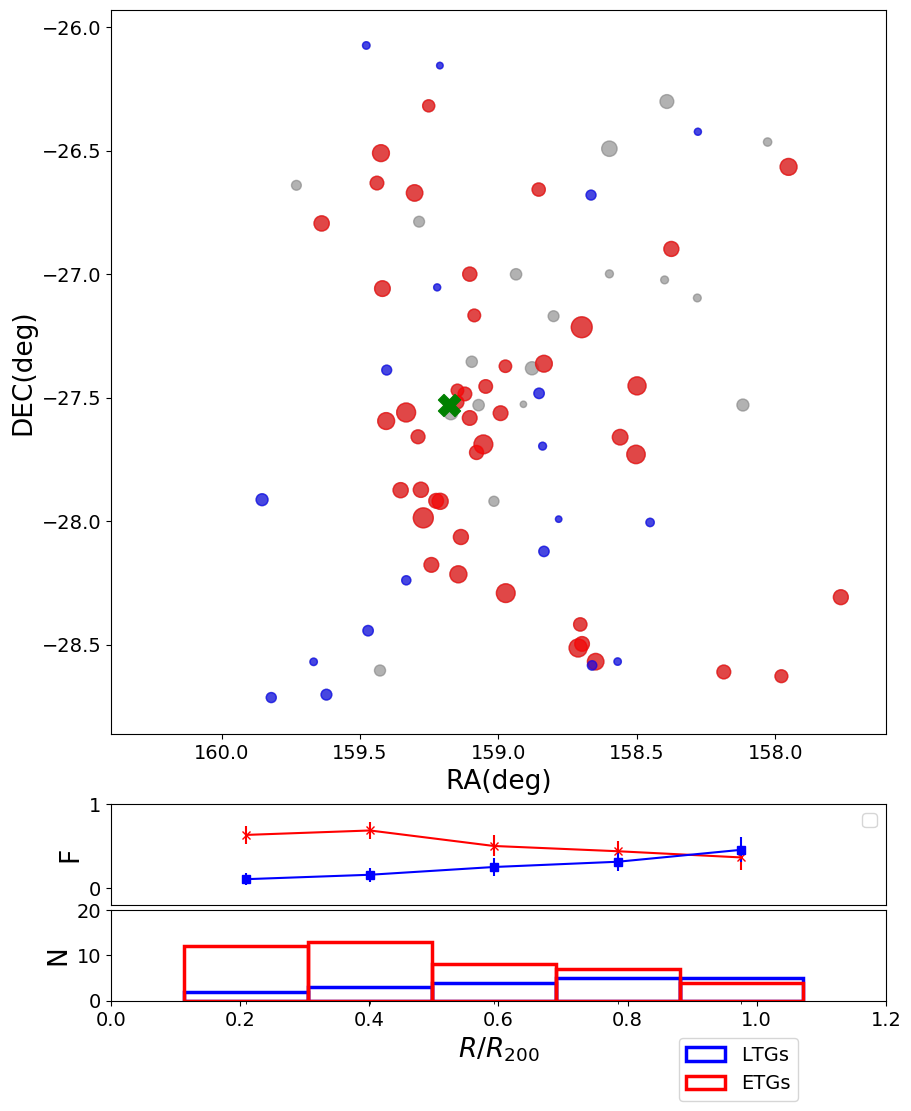}
    \caption{Top panel: Hydra's galaxies distribution. The x-axis shows the RA and y-axis shows the DEC in degrees. The radial extension is from the centre up to $\sim R_{200}$. The blue and red dots are the LTGs and ETGs respectively. The grey dots are the galaxies in the first and fourth quadrants in the Fig.~\ref{fig:early-late}. The size of each circle is proportional to the $n_{r}$. The green $x$ symbol is the cluster centre. Middle panel shows the fraction of LTGs and ETGs in blue and red respectively, as a function of distance in terms of R200. Bottom panel: histogram of galaxies per $R_{200}$ bin.}
    \label{fig:Hydra_ETG_LTG}
\end{figure}

The top panel of Fig.~\ref{fig:Hydra_ETG_LTG} shows the spatial distribution of ETGs and LTGs classified in subsection~\ref{subsec:ETG_LTG}. The red and blue dots are the ETGs and LTGs respectively, and the grey dots are the mix of galaxies that do not meet the ETG or LTG definition requirements (first and fourth quadrants in Fig.~\ref{fig:early-late}). The size of each circle is proportional to the $n_{r}$. The middle and bottom panels show the radial fraction and histograms of ETGs and LTGs, respectively. The fractions are calculated over the total number of galaxies. We find that the fraction of ETGs is higher than the LTGs fraction, basically for all bins, except at $\sim$0.9$R_{200}$, where the behaviour is inverted. We find that $\sim47$ percent of the LTGs and $\sim75$ percent of the ETGs are inside 0.6$R_{200}$. The number of LTG and ETG galaxies beyond 0.6$R_{200}$ is quite similar.

\begin{figure}
\centering

\includegraphics[width=8.0cm,height=10cm,keepaspectratio]{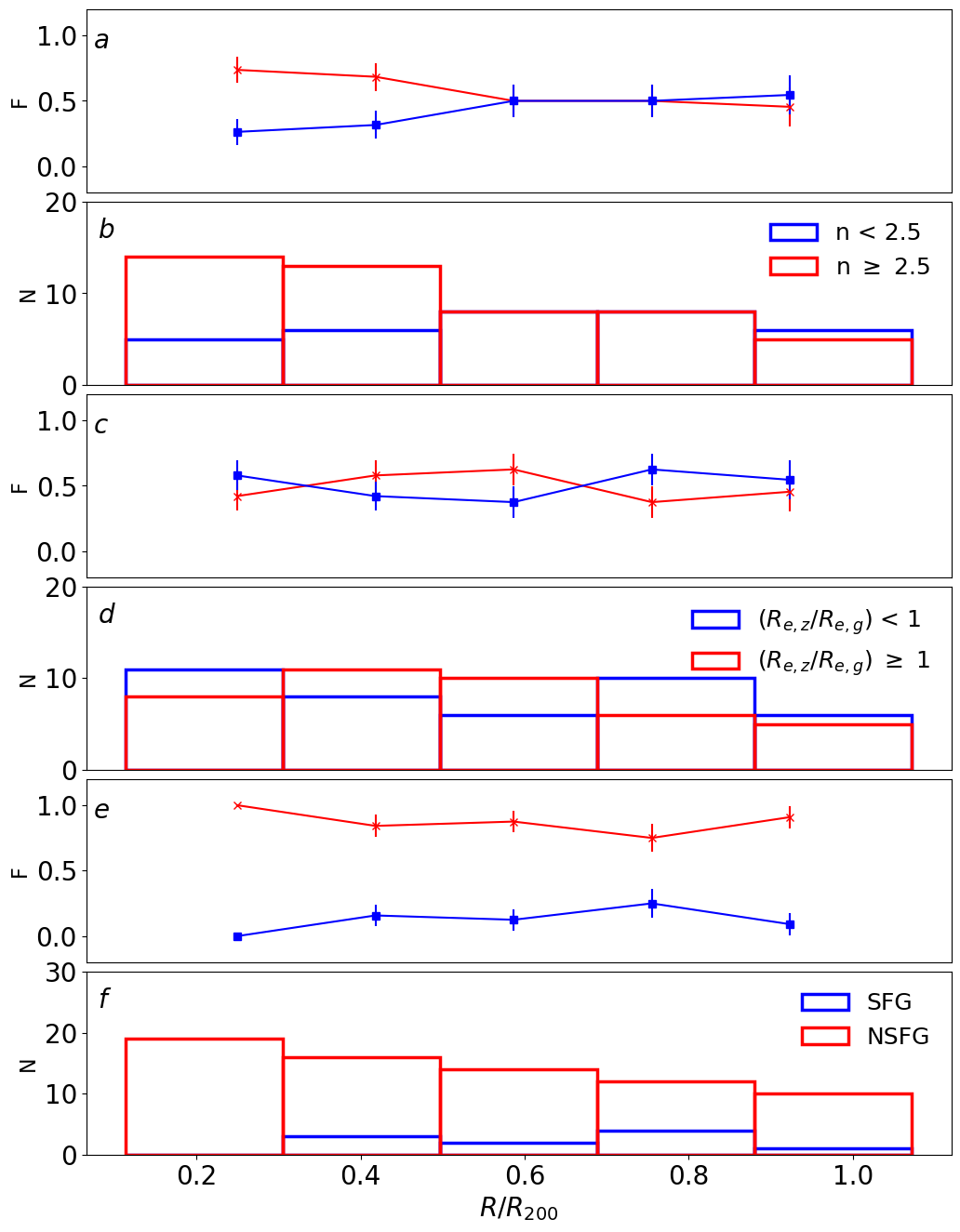}
    \caption{ The fraction (F) and histogram representing the number of galaxies (N) per $R_{200}$ bin. Panels $a$ and $b$ show galaxies with $n_{r}\geq$  2.5 and < 2.5 in red and blue respectively. Panels $c$ and $d$ show galaxies with $(R_{e,z}/R_{e,g})\geq$ 1 and < 1 in red and blue respectively, finally panels $e$ and $f$ show NSFG and SFG in red and blue respectively.}
    \label{fig:Frac_r200}
\end{figure}

Figure~\ref{fig:Frac_r200} shows the radial fraction and histograms for $n_{r}$, $(R_{e,z}/R_{e,g})$ and SFG, NSFG with respect to the clustercentric distance. Panels $a$ and $b$ show that most galaxies ($\sim59$ percent) have $n_{r}$ $\geq$ 2.5, and their fraction (panel $a$) is higher than that of galaxies with $n_{r}$ < 2.5 up to $\sim0.5R_{200}$, after that the fractions are basically the same for the two groups of galaxies. 

The $c$ and $d$ panels of Fig.~\ref{fig:Frac_r200} show the ratio of effective radii in $z$ and $g$ bands $(R_{e,z}/R_{e,g})$. Hydra has 40 galaxies with $(R_{e,z}/R_{e,g})$ > 1 and 41 otherwise, showing a similar fraction for these two cases across $R_{200}$.

Panels $e$ and $f$ of Fig~\ref{fig:Frac_r200} show the fraction and radial distribution of SFGs and NSFGs. We find that $\sim 88$ percent of galaxies in Hydra (71 galaxies) are quenched. These quenched galaxies are found at all distances from the cluster centre, however their number decreases as a function of clustercentric distance. We also find that there are no SFGs in the central bin, i.e. inside $\sim0.3R_{200}$ of Hydra. \citet{Owers2019ApJ}, who studied a sample of galaxy clusters at low redshift, also found that the number and the fraction of passive galaxies decrease towards larger clustercentric distances. They found that beyond 0.3R$_{200}$ the number of SFG remains relatively constant, while the fraction of SFGs increases as a function of clustercentric distance. \citet{Pallero2019} studied a sample of galaxy clusters from  the EAGLE hydrodynamical simulation. They classified SFGs as we do in this work, i.e. galaxies with log(sSFR) > -11, and found that a cluster as massive as Hydra (M$_{200}$ >3$\times$10$^{14}$M$_{\odot}$) has $\sim90$ percent of the galaxies quenched, in agreement with our findings. 

We show in Fig.~\ref{fig:MD_relation} the distribution of SFGs and NSFGs as a function of projected local density. No SFGs are found in the highest and lowest density bins. The NSFGs dominate at all densities, however after $\sim$ log$(\Sigma_{10} > 1.2)$ the fraction of NSFG increases towards denser regions, while the fraction of SFG decreases in the same direction. The error bars in Figures ~\ref{fig:Hydra_ETG_LTG}, ~\ref{fig:Frac_r200} and ~\ref{fig:MD_relation} are binomial uncertainties with a 68 percent confidence. 

\begin{figure}
\label{fig:MD_relation}
\includegraphics[width=8cm,height=7cm,keepaspectratio]{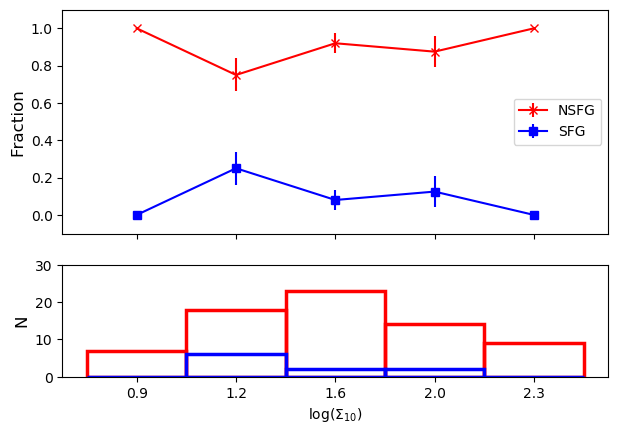}    
\caption{SF and NSF-Density relation. Top panel: fraction of SFG and NSFG in blue and red, respectively. Bottom panel: Histograms with the number of galaxies per density bin.}
    \label{fig:MD_relation}
\end{figure}

\subsection{Behaviour of $n$ as a function of the 12 S-PLUS filters}

\begin{figure}

\includegraphics[width=8.0cm,height=6cm]{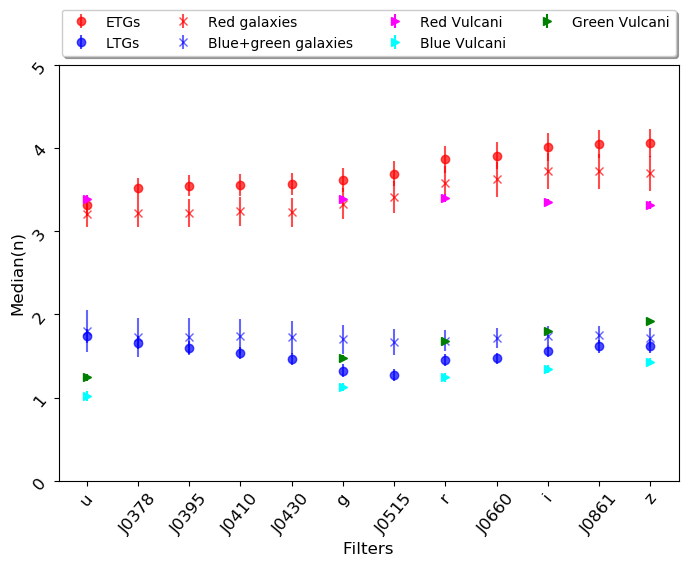}
    \caption{Median S\'ersic index $\bar{n}$ as a function of the 12 S-PLUS filters. Red symbols represent ETGs and red ($(u-r)$ $\geq$2.3) galaxies, blue symbols represent LTGs and blue + green ($(u-r)$ < 2.3) galaxies. The error bars are the standard error of the median $1.253\sigma/\sqrt{N}$, where $N$ is the number of objects. \textbf{The magenta, green and cyan symbols are the red (u-r $\geq$ 2.1), green (1.6 > (u-r) < 2.1), and blue (u-r $\leq$ 1.6) galaxies in \citet{Vulcani2014}; see Section \ref{sec:discussion} for more details. } }
    \label{fig:n_ETG_LTG_red_blue_one}
\end{figure}

The light emitted by a galaxy at different wavelengths has information from different physical phenomena. As an example, stellar populations of different ages and metallicities have emission peaks in different regions of the spectrum. In addition, other contributions such as emission from HII regions, AGN, and planetary nebulae can also be found in a galaxy spectrum. In the case of S-PLUS, each filter is placed in optical strategic regions. Thus, within this context, it is interesting to measure how $n$ changes with respect to each S-PLUS filter, clustercentric distance and density. In order to investigate it, we separate the Hydra galaxies into four groups: i) ETG, ii) LTG, iii) red and iv) blue + green, to facilitate comparisons with other studies. Galaxies with $(u-r)$ $\geq$ 2.3 are red (58 galaxies) and galaxies with $(u-r)$ < 2.3 are blue + green (23 galaxies). ETGs and LTGs follow the definition provided in Section~\ref{subsec:ETG_LTG}. Fig.~\ref{fig:n_ETG_LTG_red_blue_one} shows how the median S\'ersic index ($\bar{n}$) changes as a function of wavelength. The $\bar{n}$ for red galaxies shows a larger value (13 percent) toward redder filters. The $\bar{n}$ for blue + green galaxies remains constant across all filters. For ETGS, $\bar{n}$ increases with wavelength (18 percent), whereas for LTGs it decreases  up to the $J0515$-band, and after that increases its value up to z-band. The LTGs present a net decrease, from filter $u$ to $z$, of 7 percent. The $\bar{n}$ values and their uncertainties per filter, estimated by adding the individual uncertainties in quadrature and dividing by the number of galaxies, for ETGs, LTGs, red and blue + green galaxies are in Table~\ref{tab:n_median}.

\begin{table*}
\caption{Median of S\'ersic index ($\bar{n}$) of the galaxies per filter. The errors were estimated by adding the individual uncertainties in quadrature, divided by the number of galaxies per filter. } \label{tab:n_median}

\centering
%\scriptsize
\resizebox{\textwidth}{!}{
\begin{tabular}{lllllllllllll}
%\begin{tabular}{@{}P{0.18\textwidth}P{0.20\textwidth}P{0.12\textwidth}P{0.4\textwidth}@{}}

GALAXIES & $\bar{n}_{u}$           & $\bar{n}_{J0378}$      & $\bar{n}_{J0395}$      & $\bar{n}_{J0410}$      & $\bar{n}_{J0430}$      & $\bar{n}_{g}$          & $\bar{n}_{J0515}$      & $\bar{n}_{r}$          & $\bar{n}_{J0660}$      & $\bar{n}_{i}$          & $\bar{n}_{j0861}$      & $\bar{n}_{z}$          \\
\hline
\hline
ETG      & 3.31$\pm$0.03 & 3.52$\pm$0.02 & 3.55$\pm$0.02 & 3.56$\pm$0.02 & 3.57$\pm$0.01 & 3.62$\pm$0.01 & 3.69$\pm$0.01 & 3.87$\pm$0.01 & 3.91$\pm$0.01 & 4.02$\pm$0.01 & 4.05$\pm$0.01 & 4.06$\pm$0.01 \\
LTG      & 1.74$\pm$0.02 & 1.66$\pm$0.01 & 1.60$\pm$0.01 & 1.54$\pm$0.01 & 1.46$\pm$0.01 & 1.32$\pm$0.01 & 1.27$\pm$0.01 & 1.45$\pm$0.01 & 1.47$\pm$0.01 & 1.56$\pm$0.01 & 1.62$\pm$0.01  & 1.62$\pm$0.01 \\
RED      & 3.21$\pm$0.03 & 3.22$\pm$0.02 & 3.22$\pm$0.02 & 3.24$\pm$0.01 & 3.23$\pm$0.01 & 3.33$\pm$0.01 & 3.41$\pm$0.01 & 3.58$\pm$0.01 & 3.62$\pm$0.01 & 3.72$\pm$0.01 & 3.72$\pm$0.01  & 3.70$\pm$0.01 \\
BLUE + GREEN     & 1.80$\pm$0.02 & 1.72$\pm$0.02 & 1.73$\pm$0.01 & 1.74$\pm$0.01 & 1.72$\pm$0.01 & 1.70$\pm$0.01 & 1.67$\pm$0.01 & 1.68$\pm$0.01 & 1.71$\pm$0.01 & 1.74$\pm$0.01 & 1.75$\pm$0.01  & 1.72$\pm$0.01

\end{tabular}
}
\end{table*}

\begin{figure}

\includegraphics[width=8.0cm,height=6cm]{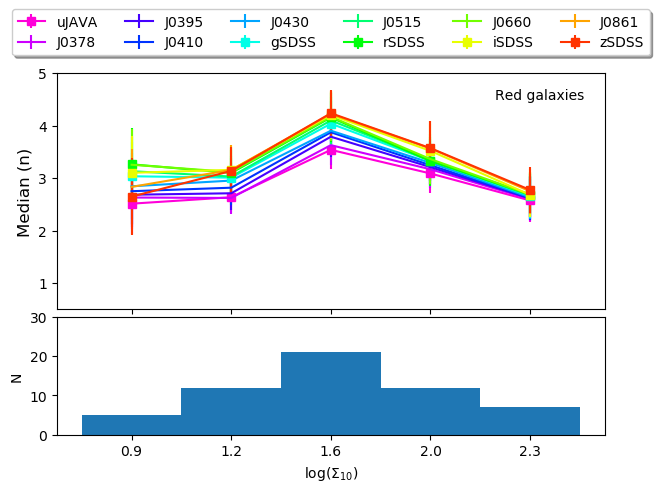}
    \caption{S\'ersic index - Density relation for red galaxies. The top panel shows the $\bar{n}$, for each S-PLUS filter, with respect to the cluster density. The bottom panel shows a histogram with the number of galaxies per density bin.  }
    \label{fig:red_D_sersic}
\end{figure}

\begin{figure}

\includegraphics[width=8cm,height=6cm]{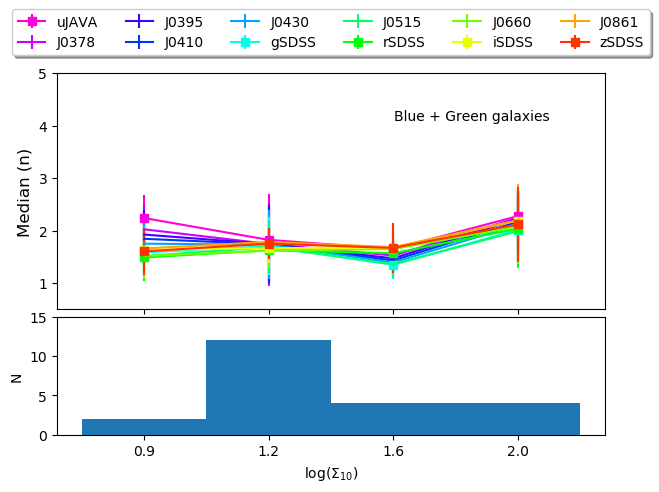}
    \caption{Same as Fig.~\ref{fig:red_D_sersic}, but for the blue + green galaxies.}
    \label{fig:blue_D_sersic}
\end{figure}

Figures~\ref{fig:red_D_sersic} and \ref{fig:blue_D_sersic} show how $\bar{n}$, for each S-PLUS filter, changes with density (quantified by $\log(\Sigma_{10})$) for red and blue + green galaxies, respectively. The $\bar{n}$ for red galaxies (Fig.~\ref{fig:red_D_sersic}), shows an increase up to $\log(\Sigma_{10})$ = 1.6, and then decreases for denser regions. For the blue + green galaxies (Fig.~\ref{fig:blue_D_sersic}) $\bar{n}$, in general, remains nearly constant at all densities. 

\begin{figure}

\includegraphics[width=8.0cm,height=6cm]{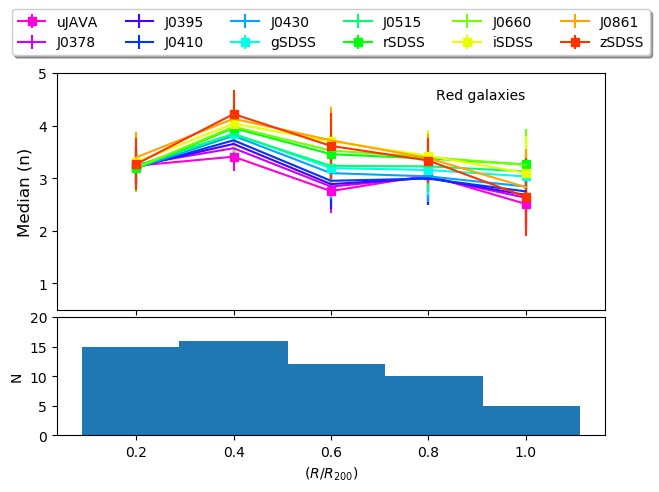}
    \caption{S\'ersic index -  Radius relation for red galaxies. The top panel shows the $\bar{n}$, for each S-PLUS filter, with respect to the cluster density. The bottom panel shows a histogram with the number of galaxies per ($R/R_{200}$) bin. }
    \label{fig:red_R_sersic}
\end{figure}

\begin{figure}

\includegraphics[width=8cm,height=6cm]{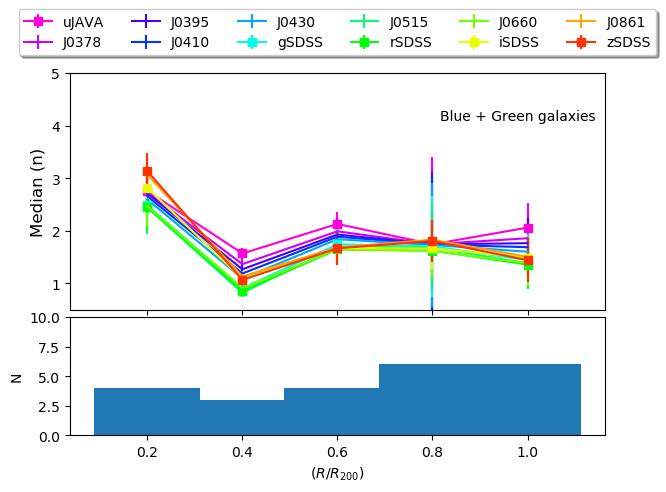}
    \caption{Same as Fig.~\ref{fig:red_R_sersic}, here for the blue + green galaxies.}
    \label{fig:blue_R_sersic}
\end{figure}

Figures~\ref{fig:red_R_sersic} and \ref{fig:blue_R_sersic} show the behaviour of $\bar{n}$ for each filter, for red and blue + green galaxies, respectively, but now as a function of the clustercentric distance. For red galaxies, Fig.~\ref{fig:red_R_sersic} shows that, beyond $\sim$0.4R$_{200}$, galaxies show lower $\bar{n}$ values, for all filters, at farther distances from the cluster centre. For blue + green galaxies, Fig.~\ref{fig:blue_R_sersic} shows that $\bar{n}$ decreases, for all filters, up to $\sim0.4R_{200}$, after that it remains constant with distance within the uncertainties. These results will be discussed in section~\ref{sec:discussion}.

\subsection{Substructures in the Hydra Cluster and the phase-space diagram}

The environment in which a galaxy is embedded may play an important role in determining its morphological (e.g. the $n$) and physical  (e.g. the sSFR) parameters. The presence of substructures in a cluster can influence the parameters mentioned above. To check if this is the case in Hydra, we need to determine whether there are possible substructures within the cluster. The substructures generally have galaxies of different brightness, thus if we only use galaxies brighter than 16 (r-band) to check for the presence of substructure, we could lose important information. Therefore, we decide to use in this section the galaxies selected in Section 2.1, i.e. all the galaxies that have peculiar velocities lower than the escape velocity of the cluster, without considering any cut in magnitudes. There is a total of 193 such galaxies in Hydra and we apply the Dressler-Schectman test \citep[DST,][]{DresslerShectman1988} on those galaxies to search for the existence of possible substructures. The DST estimates a $\Delta$ statistic for the cluster by comparing the kinematics of neighbouring galaxies with respect to the kinematics of the cluster. This comparison is done for each galaxy taking into account the $N_{nn}$ nearest neighbours, where we use $N_{nn}=\sqrt{(N_{\rm total})}$ following \citet{Pinkney1996}, with $N_{\rm total}$ being the total number of objects in the Hydra cluster. Each galaxy and their $N_{nn}$ neighbours are used to estimate a local mean velocity $\bar{v}_{local}$ and a local velocity dispersion $\sigma_{local}$, which are compared with the mean velocity ($\bar{v}$) and velocity dispersion ($\sigma$) of the cluster. For each galaxy we then estimate the iation $\delta$ as: 

\begin{equation}
   \delta^{2}=(N_{nn}/\sigma^{2})[(\bar{v}_{local}-\bar{v})^{2}+(\sigma_{local}-\sigma^{2})]
	\label{eq:DST}
\end{equation}

The $\Delta$ statistic is then defined as the cumulative deviation; $\Delta=\sum_{i}^{N}\delta_{i}$. 
If $\Delta/N_{\rm total} > 1$ it means that probably there is a substructure in the cluster \citep{White2015}. For the Hydra cluster considering an area of $\sim95'$ of radius and with 193 members, we found a $\Delta/N_{\rm total} = 1.25$. To confirm that this result is significantly different from a random distribution, and validate the possible existence of substructures in Hydra, we calibrate the $\Delta$ statistic by randomly shuffling the velocities using a Monte Carlo simulation. We perform 1x10$^5$ iterations; for each of those we calculated the $\Delta$ statistic, which we call for simplicity random $\Delta$. We then count the number of configurations for which the random $\Delta$ is greater than the original $\Delta$, N($\Delta$,randoms > $\Delta$), normalised by the number of iterations N($\Delta$,randoms), that is

\begin{equation}
 P = \frac{N(\Delta,{\rm randoms} > \Delta)}{N(\Delta,{\rm randoms})}
\end{equation}

If the fraction P is lower than 0.1, we can conclude that the original $\Delta$ value is not obtained from a random distribution \citep{White2015}. We find P = 0.016, which confirms a possible presence of substructures in the Hydra cluster.

\begin{figure}
\includegraphics[width=8cm,height=7cm,keepaspectratio]{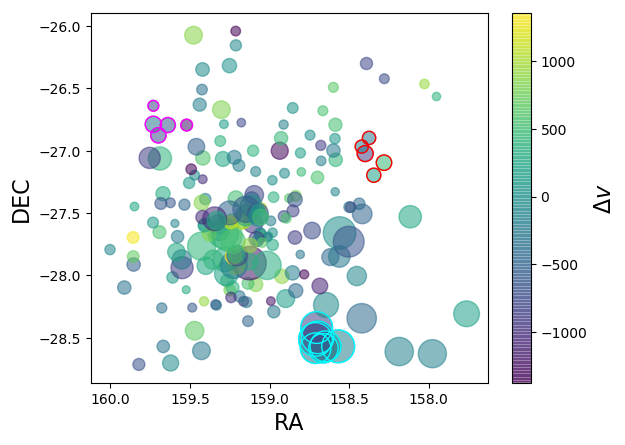}
    \caption{ Spatial distribution of Hydra galaxies. The size of each circle is proportional to $e^\delta$, i.e. bigger dots indicate that the galaxy is in a possible substructure. The galaxies are colour-coded by their peculiar velocity. The galaxies that belong to the different structures identified by {\tt DBSCAN} are enclosed by open circles of different colours.}
    \label{fig:DST}
\end{figure}

In Fig.~\ref{fig:DST} we highlight the possible substructures detected in Hydra, where we show the distribution of the cluster's galaxies where the size of each circle is proportional to $e^\delta$. Larger circles indicate a higher probability for a galaxy be part of a substructure.  Each galaxy is colour-coded by their peculiar velocities with respect to the cluster's redshift. We found possible substructures in the outer regions of the cluster as well as in the central part,these possible substructures are enclosed by a black open square in Fig~\ref{fig:PS_DST_sSFR}.

We note that \citet{Stein1997} found that Hydra does not present any substructure, by applying the same DST but in an area of 45$'$ radius (i.e half the area we use here), which includes 76 galaxies\footnote{We also note that, using the normality-test \citet{Stein1997} found that Hydra presents substructures with 1 percent of significance level, see \citet{Beers1990} for more details.}. To understand this discrepancy we applied the DST in the same area as \citet{Stein1997} using 136 galaxies, and found $\Delta/N_{\rm total} = 1.0$, which is inconclusive in terms of substructures. Therefore, we detect possible new substructures in an unexplored area of Hydra.

In order to further look into the possibility of substructures, we use a Density-Based Spatial Clustering \citep[{\tt DBSCAN},][]{ester} algorithm. {\tt DBSCAN} takes as input the positions of the objects and the minimum number of objects with a maximum distance between them to be considered as a group/substructure. We adopt a maximum distance of 140 kpc and a minimum of 3 galaxies to consider a substructure \citep{Sohn2015,Olave-Rojas2018}. Using the 193 galaxies, {\tt DBSCAN} finds the presence of three substructures within R$_{200}$, one of them in agreement with the possibility of a substructure found with the DST. Each galaxy that belongs to the substructure found by {\tt DBSCAN} is enclosed by open circles in Fig~\ref{fig:DST}. The galaxies that belong to the same structure found by DST are enclosed by cyan open circles; this substructure has 7 galaxies, 2 of them are SFGs.

\begin{figure}
\includegraphics[width=8cm,height=7cm,keepaspectratio]{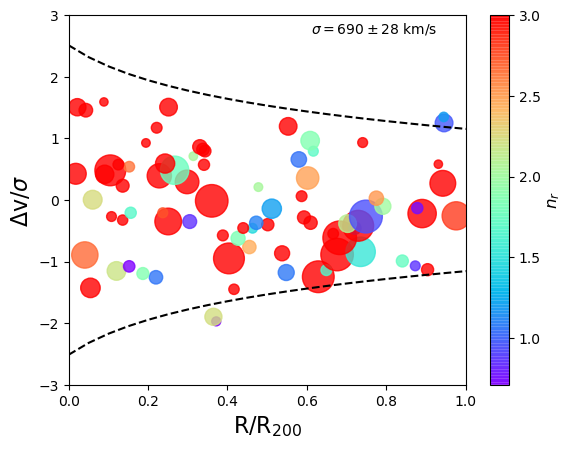}
    \caption{The phase-space diagram. The x-axis shows the projected distance from the cluster centre normalised by $R_{200}$. The y-axis shows the peculiar line-of-sight velocity with respect to the cluster recessional velocity, normalised by the velocity dispersion of the cluster. The dashed lines are the escape velocity defined in \citet[][see their Eq. 2]{Jaffe2015}. Each galaxy is colour-coded by $n_{r}$. The colour bar saturates at $n_{r}$=3, thus all galaxies that have $n_{r} \geq 2.5$ are in red. The size of each circle is proportional to the $\delta$ of the DST.}
    \label{fig:PS_DST_n}
\end{figure}

\begin{figure}
\includegraphics[width=8cm,height=7cm,keepaspectratio]{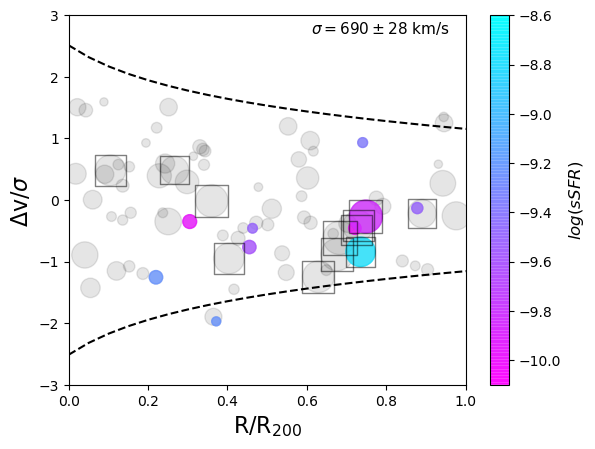}
    \caption{Same as Fig.~\ref{fig:PS_DST_n}, colour-coded by $\log(sSFR)$. The grey dots are galaxies that do not have $H\alpha$ emission. Galaxies with $\delta$> 3 times the standard deviation of $\delta$ distribution are enclosed by open black square. }
    \label{fig:PS_DST_sSFR}
\end{figure}

Having determined the possible existence of substructures in Hydra, the phase-space diagram \citep{Jaffe2015}, which relates the distance to the cluster centre with the kinematical quantity $\Delta v/\sigma$, could help us to understand the dynamic state of Hydra. We show the phase-space diagram of Hydra in Fig.~\ref{fig:PS_DST_n}, where we use the 81 galaxies that have m$_r < 16$. The $x$-axis is the projected distance from the cluster centre normalised by $R_{200}$ and the $y$-axis is the peculiar line-of-sight velocity of each galaxy with respect to the cluster recessional velocity, normalised by the velocity dispersion of the cluster. The escape velocity is indicated by the dashed lines and it is obtained  based on a Navarro, Frenk \& White dark matter profile \citep{NFW1996}, see \citet{Jaffe2015} for more details. Figs.~\ref{fig:PS_DST_n} and ~\ref{fig:PS_DST_sSFR} show the same diagram colour-coded by $n_{r}$ and $\log(sSFR)$, respectively, where the circle sizes are proportional to $\delta$. We can see that there are galaxies both with higher and low $n_{r}$ in possible substructures (bigger circles). Also, it is clear that some galaxies located in substructures are star forming. We can see from Fig.~\ref{fig:PS_DST_sSFR} that galaxies with larger $\delta$  are located beyond $0.6R_{200}$, and also some galaxies are closer to the cluster central region. Galaxies with $\delta \geq$ 3$\sigma_{\delta}$ (the standard deviation of $\delta$ distribution) have the largest probability to belonging to a substructure  \citep{Girardi1997,Olave-Rojas2018}. Fig~\ref{fig:PS_DST_sSFR} shows, enclosed by a black open square, the 12 galaxies that have $\delta \geq$ 3$\sigma_{\delta}$. Four of these galaxies are near to the cluster centre and 8 are beyond $0.6R_{200}$. Two of the 12 galaxies are SFGs and 8 have $n_{r} \geq 2.5$.

\section{discussion}\label{sec:discussion}

We discuss in this section the results that we found in this work, interpret and compare them with previous studies. Also, we discuss few caveats and cautions that should be taken into account when interpreting some of the results presented here.

We find that $\bar{n}$ for ETGs, as well as red galaxies, displays a slight increase towards redder filters (13 percent of its value for red galaxies and 18 percent for ETGs, see Fig.~\ref{fig:n_ETG_LTG_red_blue_one}). This result is in agreement with previous studies \citep{LaBarbera2010,Kelvin2012,Vulcani2014}. For LTGs, we find that $\bar{n}$ decreases, from filter $u$ to $z$, by 7 percent.  However, other studies both in cluster and in field galaxies have found that $n$ increases as a function of wavelength for LTGs \citep{LaBarbera2010,Kelvin2012,Vulcani2014,Psychogyios2019}. This behaviour may be due to higher star formation in these galaxies, which is more concentrated in the inner regions. To better compare with \citet[][hereafter V14]{Vulcani2014} we have included their data points in Fig.~\ref{fig:n_ETG_LTG_red_blue_one}. 
V14 used the filters $u$, $g$, $r$, $i$, and $z$ from \textit{SDSS}, and modelled field galaxies with a single S\'ersic profile. We note, first, that their sample contains field galaxies and, second, that their colour cuts to separate galaxies as red, blue and green are slightly different than the one used in this work. Nevertheless, the green+blue and LTGs, from Hydra, show similar values of $\bar{n}$, within the uncertainties, with respect to the green galaxies in V14 for most filters. The $\bar{n}$ value for the $u$ filter is higher for Hydra. For the red galaxies and filters $u$, $g$, and $r$ our work and that of V14 show similar values of $\bar{n}$. However, galaxies in Hydra show a higher value in the filters $i$ and $z$ when compared to the field galaxies in V14. Overall galaxies in Hydra exhibit a higher value of $\bar{n}$ when compared to the blue galaxies of V14. This result is interesting and in agreement with \citet{Psychogyios2019}, who found that cluster galaxies always have a higher value of $\bar{n}$, regardless of the filter, when compared to field galaxies. Thus, bearing in mind that a comparison with other studies is not so simple, due to the differences in separating galaxies as ETGs, LTGs, red, green and blue, which likely contribute to differences in the results, there may be actual differences between the $\bar{n}$ values as a function of filter between cluster and field galaxies. We will explore this further in a future work.

The $\bar{n}$ value for blue + green galaxies remains constant as a function of wavelength. We find that it appears to have a higher value for the bluest filters with respect to the clustercentric distance (Fig.~\ref{fig:blue_R_sersic}). However, considering the uncertainties, we find no significant change in $\bar{n}$ as a function of distance or wavelength. We can see the same behaviour with respect to the density ($\log(\Sigma_{10})$), where $\bar{n}$ remains constant, considering the uncertainties (Fig.~\ref{fig:blue_D_sersic}). The $\bar{n}$ for red galaxies is always greater than 2.5, and for blue + green galaxies is generally lower than 2. A similar result was presented in \citet{Psychogyios2019} using 5 filters (u, B, V, J and K). They found that the S\'ersic index remains nearly constant as a function of wavelength, for galaxies that belong to a cluster.

Examining the Hydra's SMF we find that the ETGs and NSFGs are more massive and dominate the higher mass end of the global SMF (see Fig. ~\ref{fig:SMF}), as expected for a cluster. The lower mass galaxies are mostly LTGs and SFGs and dominate the lower mass end of the SMF. These results are in agreement with previous studies \citep[e.g.,][]{Blanton2009ARA&,Vulcani2011MNRAS,Etherington2017MNRAS,vanderBurg2020A&A}. For example, \citet{Vulcani2011MNRAS} analysed a sample of 21 nearby clusters (0.04<z<0.07), finding that the global SMF is dominated by ETGs (Elliptical and S0 galaxies), while the number of late-type galaxies declines towards the high mass end of the SMF. In the local universe, the high mass end of the SMF is dominated  by ETGs both in low and high-dense environments \citep{Blanton2009ARA&,Bell2003}. The low mass end of the SMF for low-density environments is completely dominated by blue disk galaxies, whereas in high-dense environments there is a mix of ETGs and LTGs \citep{Blanton2009ARA&}. Regarding the red galaxies, with the increase of local density (isolated $-$ poor group $-$ rich group $-$ cluster), the characteristic mass of this population increases.  This is because most of the brightest objects are red and located in dense regions. In addition, the blue population declines in importance as the local density increases \citep{Blanton2009ARA&}.

In this work we separate galaxies as blue + green  ($(u-r)$ < 2.3  ) and red ($(u-r)\geq$2.3). There is an intermediate zone between the red sequence and the blue cloud, called green valley. These galaxies have typical colours between 1.8 < ($u-r$) <2.3 \citep{Schawinski2014MNRAS}. In Hydra, we find that galaxies with colours 1.8 < $(u-r)$  <  2.3 do not present H$\alpha$ emission. These objects could be transitional galaxies that have lower sSFRs than actively star-forming galaxies of the same mass. The galaxies in the green valley have very low levels of ongoing star formation \citep[log (sSFR) < -10.8,][]{Salin2014}, and we are not detecting star formation at such a low level. These galaxies do not meet the criteria required to be considered as emission line galaxies, i.e. to have a genuine narrow-band colour excess (the 3$\Sigma$ cut and the have EW$_{J0660}$ > 12 \AA). It is possible that these objects have $H\alpha$ emission below the detection limit of the survey. Perhaps, in some of these galaxies, deeper images will evidence $H\alpha$ emission. Most of these objects appear located in the outskirts of the cluster, which suggests that these have fallen into the cluster recently; however these have already suffered the influence of the cluster, and may already have lost some or a considerable amount of their gas, which might be another possibility to explain the absence of $H\alpha$ emission. We therefore suggest that this population of green galaxies could have stopped their star formation recently due to environmental effects \citep{Peng2010ApJ,Wright2019}.

As it is well known, the morphology and $SFR$ of galaxies are correlated, and the environment likely plays an important role in the evolutionary changes of these parameters \citep{Paulono-Afonso2019,Pallero2019,Fasano2015}. In this work we study the behaviour of sSFR with respect to the clustercentric distance and density. We find that $\sim88$ percent of the Hydra' galaxies are quenched, based on their $H\alpha$ emission.  However, 23 percent of Hydra galaxies are LTGs, according to our morphological and physical classification. This suggests that, as advocated by \citet{Liu2019}, the galactic physical properties (e.g. SFR) of galaxies in a cluster change faster than their structural properties.

The Dressler-Schectman test we performed in the previous section indicates that Hydra presents substructures with a $\Delta/N_{\rm total}$ = 1.25. This value is not too high in comparison with other clusters (see as a example \citealt{Olave-Rojas2018}). We also confirm the presence of substructures using the {\tt DBSCAN} algorithm. Interestingly, previous studies have shown that Hydra has an approximately homogeneous X-ray distribution. This observed feature suggests that the cluster has not suffered merging processes recently, and probably is a relaxed system \citep{Fitchett1988,FurushoEtAl2001,Hayakawa2004,Lokas2006}. It was also found in other work that Hydra does not present a Gaussian velocity distribution \citep{Fitchett1988}. Combining the previous information with our findings of substructures in Hydra, we conclude that Hydra is disturbed/perturbed, but not significantly and thus it is very close to become a virialized system. 

S-PLUS will observe a huge area around Hydra Cluster, which we will use to study the behaviour of the galaxies in a less dense environment, but still under the influence of the cluster (beyond $1R_{200}$). In a upcoming work, we will also study the morphological and physical parameters of the bulge and the disc components of Hydra galaxies, separately. In the case of the substructures in the Hydra cluster, deeper images will allow us to see if the substructures can be correlated to intracluster light. In addition, these observations will be compared with state-of-the-art hydrodynamical simulations such as IllUSTRIS-TNG \citep{Nelson2018} and C-EAGLE \citep{Barnes2018} in order to understand the evolutionary processes responsible of the observed properties of the Hydra cluster galaxies.

\section{Summary and Conclusions}
\label{sec:conclusions}

This is the first of a series of papers exploring the evolution of galaxies in dense environments. We used the Hydra Cluster as a laboratory to study galaxy evolution, more specifically to explore the structural and physical properties of galaxies with respect to the environment. The analysis done in this work is based on S-PLUS data, a survey that has 12 filters in the visible range of the spectrum and a camera with a field of view of $\sim$ 2 square degree. The area studied here involves four S-PLUS fields, covering approximately a region of 1.4 Mpc radius centered on the Hydra cluster. We analyse the derived structural and physical parameters of a selected sample of 81 Hydra galaxies. All of them are brighter than 16 mag ($r$-band) and are part of the cluster.

Our main findings are:

1) There is a clear correlation between $n_{r}$ and the galaxy stellar mass. Higher values of $n_{r}$ are found for larger stellar masses. The sSFR has the opposite behaviour, the higher the $n_{r}$, the lower is the sSFR, as expected (see Figures ~\ref{fig:early-late} and \ref{fig:early-late-ssfr}).

2) There is a larger fraction of NSFGs (galaxies with $\log(sSFR)\leq$ -11), than SFGs (galaxies with $\log(sSFR)$ > -11) across all R$_{200}$ and all cluster densities. We find that $\sim$88 percent of Hydra galaxies can be classified as NSFG.

3) The $\bar{n}$ changes with the colour of the galaxies. The $\bar{n}$ for red and ETGs present an increase of 13 and 18 percent respectively, towards redder wavelengths. The LTGs shows a decrease in $\bar{n}$ (7 percent), from $u$ to $z$ band, while the $\bar{n}$ for blue + green galaxies remains constant. Beyond $\sim$0.3R$_{200}$ red galaxies show lower $\bar{n}$ values, for all filters, declining as a function of distance from the cluster centre. The $\bar{n}$ for blue + green galaxies remains constant for all filters.

4) We find that the Hydra cluster presents possible substructures, as determined from a Dressler-Schectman test (DST). The Density-Based Spatial Clustering {\tt DBSCAN} algorithm found a substructure with exactly the same galaxies as one of the substructures detected by the DST. Two of the galaxies that are in that likely substructure are SFG, and lie in front of the cluster centre, based on the phase-space diagram. We speculate that these galaxies are falling now in to the cluster. However, given that there are few substructure, we conclude that the Hydra cluster is, although perturbed, close to virialization.

\section*{Acknowledgements}
We thank the anonymous referee for the useful comments that helped us to improve this paper. We would like to thank the S-PLUS for the early release of the data, that made this study possible. CL-D acknowledges scholarship from CONICYT-PFCHA/Doctorado Nacional/2019-21191938. CL-D and AM acknowledge support from FONDECYT Regular grant 1181797. CL-D  acknowledges also the support given by the `Vicerrector\'ia de Investigacion de la Univesidad de La Serena' program `Apoyo al fortalecimiento de grupos de investigacion'.CL-D and AC acknowledges to Steven Bamford and Boris Haeussler with the MegaMorph project. CL-D and DP acknowledge support from fellowship `Becas Doctorales Institucionales ULS', granted by the `Vicerrector\'ia de Investigacion y Postgrado de la Universidad de La Serena'. AM and DP acknowledge funding from the Max Planck Society through a “Partner Group” grant. DP acknowledges support from FONDECYT Regular grant Nr. 1181264. This work has made use of the computing facilities of the Laboratory of Astroinformatics (Instituto de Astronomia, Geof\'isica e Ci\^encias Atmosf\'ericas, Departamento de Astronomia/USP, NAT/Unicsul), whose purchase was made possible by FAPESP (grant 2009/54006-4) and the INCT-A. Y.J. acknowledges financial support from CONICYT PAI (Concurso Nacional de Inserci\'on en la Academia 2017) No. 79170132 and FONDECYT Iniciaci\'on 2018 No. 11180558. LS thanks the FAPESP scholarship grant 2016/21664-2. AAC acknowledges support from FAPERJ (grant E26/203.186/2016), CNPq (grants  304971/2016-2  and 401669/2016-5), and the Universidad de Alicante (contract UATALENTO18-02). AMB thanks the FAPESP scholarship grant 2014/11806-9. RA acknowedges support from ANID FONDECYT Regular Grant 1202007.

\section*{Data Availability}

The data used in this article is from an internal release of the S-PLUS. This means that the sample studied here is not publicly available now but will be included in the next public data release of S-PLUS.

%%%%%%%%%%%%%%%%%%%%%%%%%%%%%%%%%%%%%%%%%%%%%%%%%%



\bibliographystyle{mnras}
\bibliography{hydra} % if your bibtex file is called example.bib

\begin{thebibliography}{}
\makeatletter
\relax
\def\mn@urlcharsother{\let\do\@makeother \do\$\do\&\do\#\do\^\do\_\do\%\do\~}
\def\mn@doi{\begingroup\mn@urlcharsother \@ifnextchar [ {\mn@doi@}
  {\mn@doi@[]}}
\def\mn@doi@[#1]#2{\def\@tempa{#1}\ifx\@tempa\@empty \href
  {http://dx.doi.org/#2} {doi:#2}\else \href {http://dx.doi.org/#2} {#1}\fi
  \endgroup}
\def\mn@eprint#1#2{\mn@eprint@#1:#2::\@nil}
\def\mn@eprint@arXiv#1{\href {http://arxiv.org/abs/#1} {{\tt arXiv:#1}}}
\def\mn@eprint@dblp#1{\href {http://dblp.uni-trier.de/rec/bibtex/#1.xml}
  {dblp:#1}}
\def\mn@eprint@#1:#2:#3:#4\@nil{\def\@tempa {#1}\def\@tempb {#2}\def\@tempc
  {#3}\ifx \@tempc \@empty \let \@tempc \@tempb \let \@tempb \@tempa \fi \ifx
  \@tempb \@empty \def\@tempb {arXiv}\fi \@ifundefined
  {mn@eprint@\@tempb}{\@tempb:\@tempc}{\expandafter \expandafter \csname
  mn@eprint@\@tempb\endcsname \expandafter{\@tempc}}}

\bibitem[\protect\citeauthoryear{{Abadi}, {Moore}  \& {Bower}}{{Abadi}
  et~al.}{1999}]{Abadi1999}
{Abadi} M.~G.,  {Moore} B.,   {Bower} R.~G.,  1999, \mn@doi [\mnras]
  {10.1046/j.1365-8711.1999.02715.x}, \href
  {https://ui.adsabs.harvard.edu/abs/1999MNRAS.308..947A} {308, 947}

\bibitem[\protect\citeauthoryear{{Abazajian} et~al.,}{{Abazajian}
  et~al.}{2009}]{Abazajian2009}
{Abazajian} K.~N.,  et~al., 2009, \mn@doi [\apjs]
  {10.1088/0067-0049/182/2/543}, \href
  {https://ui.adsabs.harvard.edu/abs/2009ApJS..182..543A} {182, 543}

\bibitem[\protect\citeauthoryear{{Arnaboldi}, {Ventimiglia}, {Iodice},
  {Gerhard}  \& {Coccato}}{{Arnaboldi} et~al.}{2012}]{Arnaboldi2012}
{Arnaboldi} M.,  {Ventimiglia} G.,  {Iodice} E.,  {Gerhard} O.,   {Coccato} L.,
   2012, \mn@doi [\aap] {10.1051/0004-6361/201116752}, \href
  {http://adsabs.harvard.edu/abs/2012A%26A...545A..37A} {545, A37}

\bibitem[\protect\citeauthoryear{{Arnouts}, {Cristiani}, {Moscardini},
  {Matarrese}, {Lucchin}, {Fontana}  \& {Giallongo}}{{Arnouts}
  et~al.}{1999}]{Arnouts1999}
{Arnouts} S.,  {Cristiani} S.,  {Moscardini} L.,  {Matarrese} S.,  {Lucchin}
  F.,  {Fontana} A.,   {Giallongo} E.,  1999, \mn@doi [\mnras]
  {10.1046/j.1365-8711.1999.02978.x}, \href
  {http://adsabs.harvard.edu/abs/1999MNRAS.310..540A} {310, 540}

\bibitem[\protect\citeauthoryear{{Babyk} \& {Vavilova}}{{Babyk} \&
  {Vavilova}}{2013}]{BabykEtAl2013}
{Babyk} I.~V.,  {Vavilova} I.~B.,  2013, Odessa Astronomical Publications,
  \href {http://adsabs.harvard.edu/abs/2013OAP....26..175B} {26, 175}

\bibitem[\protect\citeauthoryear{{Balogh} \& {Morris}}{{Balogh} \&
  {Morris}}{2000}]{Balogh2000}
{Balogh} M.~L.,  {Morris} S.~L.,  2000, \mn@doi [\mnras]
  {10.1046/j.1365-8711.2000.03826.x}, \href
  {https://ui.adsabs.harvard.edu/abs/2000MNRAS.318..703B} {318, 703}

\bibitem[\protect\citeauthoryear{{Bamford}, {H{\"a}u{\ss}ler}, {Rojas}  \&
  {Borch}}{{Bamford} et~al.}{2011}]{Bamford2011}
{Bamford} S.~P.,  {H{\"a}u{\ss}ler} B.,  {Rojas} A.,   {Borch} A.,  2011, in
  {Evans} I.~N.,  {Accomazzi} A.,  {Mink} D.~J.,   {Rots} A.~H.,  eds,
  Astronomical Society of the Pacific Conference Series Vol. 442, Astronomical
  Data Analysis Software and Systems XX. p.~479

\bibitem[\protect\citeauthoryear{{Barbosa}, {Arnaboldi}, {Coccato}, {Gerhard},
  {Mendes de Oliveira}, {Hilker}  \& {Richtler}}{{Barbosa}
  et~al.}{2018}]{Barbosa2018A&A}
{Barbosa} C.~E.,  {Arnaboldi} M.,  {Coccato} L.,  {Gerhard} O.,  {Mendes de
  Oliveira} C.,  {Hilker} M.,   {Richtler} T.,  2018, \mn@doi [\aap]
  {10.1051/0004-6361/201731834}, \href
  {https://ui.adsabs.harvard.edu/abs/2018A&A...609A..78B} {609, A78}

\bibitem[\protect\citeauthoryear{{Barnes} et~al.,}{{Barnes}
  et~al.}{2018}]{Barnes2018}
{Barnes} L.~A.,  et~al., 2018, \mn@doi [\mnras] {10.1093/mnras/sty846}, \href
  {https://ui.adsabs.harvard.edu/abs/2018MNRAS.477.3727B} {477, 3727}

\bibitem[\protect\citeauthoryear{{Bautz} \& {Morgan}}{{Bautz} \&
  {Morgan}}{1970}]{BautzMorgan1970}
{Bautz} L.~P.,  {Morgan} W.~W.,  1970, \mn@doi [\apjl] {10.1086/180643}, \href
  {http://adsabs.harvard.edu/abs/1970ApJ...162L.149B} {162, L149}

\bibitem[\protect\citeauthoryear{{Beers}, {Flynn}  \& {Gebhardt}}{{Beers}
  et~al.}{1990a}]{Beers1990AJ}
{Beers} T.~C.,  {Flynn} K.,   {Gebhardt} K.,  1990a, \mn@doi [\aj]
  {10.1086/115487}, \href
  {https://ui.adsabs.harvard.edu/abs/1990AJ....100...32B} {100, 32}

\bibitem[\protect\citeauthoryear{{Beers}, {Flynn}  \& {Gebhardt}}{{Beers}
  et~al.}{1990b}]{Beers1990}
{Beers} T.~C.,  {Flynn} K.,   {Gebhardt} K.,  1990b, \mn@doi [\aj]
  {10.1086/115487}, \href
  {https://ui.adsabs.harvard.edu/abs/1990AJ....100...32B} {100, 32}

\bibitem[\protect\citeauthoryear{{Bell}, {McIntosh}, {Katz}  \&
  {Weinberg}}{{Bell} et~al.}{2003}]{Bell2003}
{Bell} E.~F.,  {McIntosh} D.~H.,  {Katz} N.,   {Weinberg} M.~D.,  2003, \mn@doi
  [\apjs] {10.1086/378847}, \href
  {http://adsabs.harvard.edu/abs/2003ApJS..149..289B} {149, 289}

\bibitem[\protect\citeauthoryear{{Bell} et~al.,}{{Bell}
  et~al.}{2004}]{Bell2004ApJ}
{Bell} E.~F.,  et~al., 2004, \mn@doi [\apj] {10.1086/420778}, \href
  {http://adsabs.harvard.edu/abs/2004ApJ...608..752B} {608, 752}

\bibitem[\protect\citeauthoryear{{Bertin} \& {Arnouts}}{{Bertin} \&
  {Arnouts}}{1996}]{Bertin1996}
{Bertin} E.,  {Arnouts} S.,  1996, \mn@doi [\aaps] {10.1051/aas:1996164}, \href
  {https://ui.adsabs.harvard.edu/abs/1996A&AS..117..393B} {117, 393}

\bibitem[\protect\citeauthoryear{{Blanton} \& {Moustakas}}{{Blanton} \&
  {Moustakas}}{2009}]{Blanton2009ARA&}
{Blanton} M.~R.,  {Moustakas} J.,  2009, \mn@doi [\araa]
  {10.1146/annurev-astro-082708-101734}, \href
  {https://ui.adsabs.harvard.edu/abs/2009ARA&A..47..159B} {47, 159}

\bibitem[\protect\citeauthoryear{{Boselli} et~al.,}{{Boselli}
  et~al.}{2005}]{Boselli05}
{Boselli} A.,  et~al., 2005, \mn@doi [\apjl] {10.1086/444534}, \href
  {https://ui.adsabs.harvard.edu/abs/2005ApJ...629L..29B} {629, L29}

\bibitem[\protect\citeauthoryear{{Bruzual} \& {Charlot}}{{Bruzual} \&
  {Charlot}}{2003}]{BruzualCharlot2003}
{Bruzual} G.,  {Charlot} S.,  2003, \mn@doi [\mnras]
  {10.1046/j.1365-8711.2003.06897.x}, \href
  {http://adsabs.harvard.edu/abs/2003MNRAS.344.1000B} {344, 1000}

\bibitem[\protect\citeauthoryear{{Bruzual A.} \& {Charlot}}{{Bruzual A.} \&
  {Charlot}}{1993}]{Bruzual1993}
{Bruzual A.} G.,  {Charlot} S.,  1993, \mn@doi [\apj] {10.1086/172385}, \href
  {http://adsabs.harvard.edu/abs/1993ApJ...405..538B} {405, 538}

\bibitem[\protect\citeauthoryear{{Butcher} \& {Oemler}}{{Butcher} \&
  {Oemler}}{1978}]{Butcher78}
{Butcher} H.,  {Oemler} A. J.,  1978, \mn@doi [\apj] {10.1086/156640}, \href
  {https://ui.adsabs.harvard.edu/abs/1978ApJ...226..559B} {226, 559}

\bibitem[\protect\citeauthoryear{{Calzetti}, {Armus}, {Bohlin}, {Kinney},
  {Koornneef}  \& {Storchi-Bergmann}}{{Calzetti} et~al.}{2000}]{Calzetti2000}
{Calzetti} D.,  {Armus} L.,  {Bohlin} R.~C.,  {Kinney} A.~L.,  {Koornneef} J.,
   {Storchi-Bergmann} T.,  2000, \mn@doi [\apj] {10.1086/308692}, \href
  {http://adsabs.harvard.edu/abs/2000ApJ...533..682C} {533, 682}

\bibitem[\protect\citeauthoryear{{Cantalupo}}{{Cantalupo}}{2010}]{Cantalupo10}
{Cantalupo} S.,  2010, \mn@doi [\mnras] {10.1111/j.1745-3933.2010.00806.x},
  \href {https://ui.adsabs.harvard.edu/abs/2010MNRAS.403L..16C} {403, L16}

\bibitem[\protect\citeauthoryear{{Cardelli}, {Clayton}  \& {Mathis}}{{Cardelli}
  et~al.}{1989}]{CCM1989}
{Cardelli} J.~A.,  {Clayton} G.~C.,   {Mathis} J.~S.,  1989, \mn@doi [\apj]
  {10.1086/167900}, \href
  {https://ui.adsabs.harvard.edu/abs/1989ApJ...345..245C} {345, 245}

\bibitem[\protect\citeauthoryear{{Cenarro} et~al.,}{{Cenarro}
  et~al.}{2019}]{Cenarro2019}
{Cenarro} A.~J.,  et~al., 2019, \mn@doi [\aap] {10.1051/0004-6361/201833036},
  \href {https://ui.adsabs.harvard.edu/abs/2019A&A...622A.176C} {622, A176}

\bibitem[\protect\citeauthoryear{{Chabrier}}{{Chabrier}}{2003}]{Chabrier2003}
{Chabrier} G.,  2003, \mn@doi [\pasp] {10.1086/376392}, \href
  {http://adsabs.harvard.edu/abs/2003PASP..115..763C} {115, 763}

\bibitem[\protect\citeauthoryear{{Cicone} et~al.,}{{Cicone}
  et~al.}{2014}]{Cicone14}
{Cicone} C.,  et~al., 2014, \mn@doi [\aap] {10.1051/0004-6361/201322464}, \href
  {https://ui.adsabs.harvard.edu/abs/2014A&A...562A..21C} {562, A21}

\bibitem[\protect\citeauthoryear{{Cid Fernandes}, {Mateus}, {Sodr{\'e}},
  {Stasi{\'n}ska}  \& {Gomes}}{{Cid Fernandes} et~al.}{2005}]{Cid2005MNRAS}
{Cid Fernandes} R.,  {Mateus} A.,  {Sodr{\'e}} L.,  {Stasi{\'n}ska} G.,
  {Gomes} J.~M.,  2005, \mn@doi [\mnras] {10.1111/j.1365-2966.2005.08752.x},
  \href {https://ui.adsabs.harvard.edu/abs/2005MNRAS.358..363C} {358, 363}

\bibitem[\protect\citeauthoryear{{Coe} et~al.,}{{Coe}
  et~al.}{2012}]{Coe2012ApJ...757...22C}
{Coe} D.,  et~al., 2012, \mn@doi [\apj] {10.1088/0004-637X/757/1/22}, \href
  {https://ui.adsabs.harvard.edu/abs/2012ApJ...757...22C} {757, 22}

\bibitem[\protect\citeauthoryear{{Comerford} \& {Natarajan}}{{Comerford} \&
  {Natarajan}}{2007}]{Comerford2007}
{Comerford} J.~M.,  {Natarajan} P.,  2007, \mn@doi [\mnras]
  {10.1111/j.1365-2966.2007.11934.x}, \href
  {http://adsabs.harvard.edu/abs/2007MNRAS.379..190C} {379, 190}

\bibitem[\protect\citeauthoryear{{Cora}, {Hough}, {Vega-Mart{\'\i}nez}  \&
  {Orsi}}{{Cora} et~al.}{2019}]{Cora2019}
{Cora} S.~A.,  {Hough} T.,  {Vega-Mart{\'\i}nez} C.~A.,   {Orsi} A.~A.,  2019,
  Boletin de la Asociacion Argentina de Astronomia La Plata Argentina, \href
  {https://ui.adsabs.harvard.edu/abs/2019BAAA...61..204C} {61, 204}

\bibitem[\protect\citeauthoryear{{Croton} et~al.,}{{Croton}
  et~al.}{2006}]{Croton06}
{Croton} D.~J.,  et~al., 2006, \mn@doi [\mnras]
  {10.1111/j.1365-2966.2005.09675.x}, \href
  {https://ui.adsabs.harvard.edu/abs/2006MNRAS.365...11C} {365, 11}

\bibitem[\protect\citeauthoryear{{Davis} \& {Geller}}{{Davis} \&
  {Geller}}{1976}]{Davis76}
{Davis} M.,  {Geller} M.~J.,  1976, \mn@doi [\apj] {10.1086/154575}, \href
  {https://ui.adsabs.harvard.edu/abs/1976ApJ...208...13D} {208, 13}

\bibitem[\protect\citeauthoryear{{Dekel} \& {Silk}}{{Dekel} \&
  {Silk}}{1986}]{Dekel86}
{Dekel} A.,  {Silk} J.,  1986, \mn@doi [\apj] {10.1086/164050}, \href
  {https://ui.adsabs.harvard.edu/abs/1986ApJ...303...39D} {303, 39}

\bibitem[\protect\citeauthoryear{{Desai} et~al.,}{{Desai}
  et~al.}{2007}]{Desai07}
{Desai} V.,  et~al., 2007, \mn@doi [\apj] {10.1086/513310}, \href
  {https://ui.adsabs.harvard.edu/abs/2007ApJ...660.1151D} {660, 1151}

\bibitem[\protect\citeauthoryear{{Diaferio}}{{Diaferio}}{1999}]{Diaferio1999}
{Diaferio} A.,  1999, \mn@doi [\mnras] {10.1046/j.1365-8711.1999.02864.x},
  \href {http://adsabs.harvard.edu/abs/1999MNRAS.309..610D} {309, 610}

\bibitem[\protect\citeauthoryear{{Dimauro} et~al.,}{{Dimauro}
  et~al.}{2018}]{Dimauro2018MNRAS}
{Dimauro} P.,  et~al., 2018, \mn@doi [\mnras] {10.1093/mnras/sty1379}, \href
  {https://ui.adsabs.harvard.edu/abs/2018MNRAS.478.5410D} {478, 5410}

\bibitem[\protect\citeauthoryear{{Dobrycheva}, {Vavilova}, {Melnyk}  \&
  {Elyiv}}{{Dobrycheva} et~al.}{2017}]{Dobrycheva2017}
{Dobrycheva} D.~V.,  {Vavilova} I.~B.,  {Melnyk} O.~V.,   {Elyiv} A.~A.,  2017,
  preprint, \href {http://adsabs.harvard.edu/abs/2017arXiv171208955D} {}
  (\mn@eprint {arXiv} {1712.08955})

\bibitem[\protect\citeauthoryear{{Dressler}}{{Dressler}}{1980}]{Dressler1980}
{Dressler} A.,  1980, \mn@doi [\apj] {10.1086/157753}, \href
  {http://adsabs.harvard.edu/abs/1980ApJ...236..351D} {236, 351}

\bibitem[\protect\citeauthoryear{{Dressler} \& {Shectman}}{{Dressler} \&
  {Shectman}}{1988}]{DresslerShectman1988}
{Dressler} A.,  {Shectman} S.~A.,  1988, \mn@doi [\aj] {10.1086/114694}, \href
  {https://ui.adsabs.harvard.edu/abs/1988AJ.....95..985D} {95, 985}

\bibitem[\protect\citeauthoryear{{Dressler} et~al.,}{{Dressler}
  et~al.}{1997}]{Dressler97}
{Dressler} A.,  et~al., 1997, \mn@doi [\apj] {10.1086/304890}, \href
  {https://ui.adsabs.harvard.edu/abs/1997ApJ...490..577D} {490, 577}

\bibitem[\protect\citeauthoryear{{Duc} \& {Bournaud}}{{Duc} \&
  {Bournaud}}{2008}]{Duc08}
{Duc} P.-A.,  {Bournaud} F.,  2008, \mn@doi [\apj] {10.1086/524868}, \href
  {https://ui.adsabs.harvard.edu/abs/2008ApJ...673..787D} {673, 787}

\bibitem[\protect\citeauthoryear{{Efstathiou}}{{Efstathiou}}{2000}]{Efstathiou00}
{Efstathiou} G.,  2000, \mn@doi [\mnras] {10.1046/j.1365-8711.2000.03665.x},
  \href {https://ui.adsabs.harvard.edu/abs/2000MNRAS.317..697E} {317, 697}

\bibitem[\protect\citeauthoryear{Ester, Kriegel, Sander  \& Xu}{Ester
  et~al.}{1996}]{ester}
Ester M.,  Kriegel H.-P.,  Sander J.,   Xu X.,  1996, in Proceedings of the
  Second International Conference on Knowledge Discovery and Data Mining.
  KDD'96.
AAAI Press, pp 226--231, \url
  {http://dl.acm.org/citation.cfm?id=3001460.3001507}

\bibitem[\protect\citeauthoryear{{Etherington} et~al.,}{{Etherington}
  et~al.}{2017}]{Etherington2017MNRAS}
{Etherington} J.,  et~al., 2017, \mn@doi [\mnras] {10.1093/mnras/stw3069},
  \href {https://ui.adsabs.harvard.edu/abs/2017MNRAS.466..228E} {466, 228}

\bibitem[\protect\citeauthoryear{{Faber} \& {Gallagher}}{{Faber} \&
  {Gallagher}}{1979}]{Faber1979ARA&A}
{Faber} S.~M.,  {Gallagher} J.~S.,  1979, \mn@doi [\araa]
  {10.1146/annurev.aa.17.090179.001031}, \href
  {https://ui.adsabs.harvard.edu/abs/1979ARA&A..17..135F} {17, 135}

\bibitem[\protect\citeauthoryear{{Fabian}}{{Fabian}}{2012}]{Fabian2012ARA&A}
{Fabian} A.~C.,  2012, \mn@doi [\araa] {10.1146/annurev-astro-081811-125521},
  \href {https://ui.adsabs.harvard.edu/abs/2012ARA&A..50..455F} {50, 455}

\bibitem[\protect\citeauthoryear{{Fasano}, {Poggianti}, {Couch}, {Bettoni},
  {Kj{\ae}rgaard}  \& {Moles}}{{Fasano} et~al.}{2000}]{Fasano00}
{Fasano} G.,  {Poggianti} B.~M.,  {Couch} W.~J.,  {Bettoni} D.,
  {Kj{\ae}rgaard} P.,   {Moles} M.,  2000, \mn@doi [\apj] {10.1086/317047},
  \href {https://ui.adsabs.harvard.edu/abs/2000ApJ...542..673F} {542, 673}

\bibitem[\protect\citeauthoryear{{Fasano} et~al.,}{{Fasano}
  et~al.}{2015}]{Fasano2015}
{Fasano} G.,  et~al., 2015, \mn@doi [\mnras] {10.1093/mnras/stv500}, \href
  {http://adsabs.harvard.edu/abs/2015MNRAS.449.3927F} {449, 3927}

\bibitem[\protect\citeauthoryear{{Fitchett} \& {Merritt}}{{Fitchett} \&
  {Merritt}}{1988}]{Fitchett1988}
{Fitchett} M.,  {Merritt} D.,  1988, \mn@doi [\apj] {10.1086/166902}, \href
  {http://adsabs.harvard.edu/abs/1988ApJ...335...18F} {335, 18}

\bibitem[\protect\citeauthoryear{{Furusho} et~al.,}{{Furusho}
  et~al.}{2001}]{FurushoEtAl2001}
{Furusho} T.,  et~al., 2001, \mn@doi [\pasj] {10.1093/pasj/53.3.421}, \href
  {http://adsabs.harvard.edu/abs/2001PASJ...53..421F} {53, 421}

\bibitem[\protect\citeauthoryear{{Gil de Paz} et~al.,}{{Gil de Paz}
  et~al.}{2007}]{GildePaz2007}
{Gil de Paz} A.,  et~al., 2007, \mn@doi [\apjs] {10.1086/516636}, \href
  {http://adsabs.harvard.edu/abs/2007ApJS..173..185G} {173, 185}

\bibitem[\protect\citeauthoryear{{Girardi}, {Escalera}, {Fadda}, {Giuricin},
  {Mardirossian}  \& {Mezzetti}}{{Girardi} et~al.}{1997}]{Girardi1997}
{Girardi} M.,  {Escalera} E.,  {Fadda} D.,  {Giuricin} G.,  {Mardirossian} F.,
   {Mezzetti} M.,  1997, \mn@doi [\apj] {10.1086/304113}, \href
  {https://ui.adsabs.harvard.edu/abs/1997ApJ...482...41G} {482, 41}

\bibitem[\protect\citeauthoryear{{Gonzalez} et~al.,}{{Gonzalez}
  et~al.}{2018}]{Gonzalez2018}
{Gonzalez} E.~J.,  et~al., 2018, \mn@doi [\aap] {10.1051/0004-6361/201732003},
  \href {https://ui.adsabs.harvard.edu/abs/2018A&A...611A..78G} {611, A78}

\bibitem[\protect\citeauthoryear{{Gunn} \& {Gott}}{{Gunn} \&
  {Gott}}{1972}]{Gunn1972}
{Gunn} J.~E.,  {Gott} J.~Richard I.,  1972, \mn@doi [\apj] {10.1086/151605},
  \href {https://ui.adsabs.harvard.edu/abs/1972ApJ...176....1G} {176, 1}

\bibitem[\protect\citeauthoryear{{Harrison}}{{Harrison}}{1974}]{Harrison1974ApJ}
{Harrison} E.~R.,  1974, \mn@doi [\apjl] {10.1086/181545}, \href
  {https://ui.adsabs.harvard.edu/abs/1974ApJ...191L..51H} {191, L51}

\bibitem[\protect\citeauthoryear{{Hatfield} \& {Jarvis}}{{Hatfield} \&
  {Jarvis}}{2017}]{Hatfield2017}
{Hatfield} P.~W.,  {Jarvis} M.~J.,  2017, \mn@doi [\mnras]
  {10.1093/mnras/stx2155}, \href
  {http://adsabs.harvard.edu/abs/2017MNRAS.472.3570H} {472, 3570}

\bibitem[\protect\citeauthoryear{{H{\"a}u{\ss}ler} et~al.,}{{H{\"a}u{\ss}ler}
  et~al.}{2013}]{Haubler2013}
{H{\"a}u{\ss}ler} B.,  et~al., 2013, \mn@doi [\mnras] {10.1093/mnras/sts633},
  \href {http://adsabs.harvard.edu/abs/2013MNRAS.430..330H} {430, 330}

\bibitem[\protect\citeauthoryear{{Hayakawa}, {Furusho}, {Yamasaki}, {Ishida}
  \& {Ohashi}}{{Hayakawa} et~al.}{2004}]{Hayakawa2004}
{Hayakawa} A.,  {Furusho} T.,  {Yamasaki} N.~Y.,  {Ishida} M.,   {Ohashi} T.,
  2004, \mn@doi [\pasj] {10.1093/pasj/56.5.743}, \href
  {http://adsabs.harvard.edu/abs/2004PASJ...56..743H} {56, 743}

\bibitem[\protect\citeauthoryear{{Hopkins}, {Connolly}, {Haarsma}  \&
  {Cram}}{{Hopkins} et~al.}{2001}]{Hopkins2001AJ}
{Hopkins} A.~M.,  {Connolly} A.~J.,  {Haarsma} D.~B.,   {Cram} L.~E.,  2001,
  \mn@doi [\aj] {10.1086/321113}, \href
  {https://ui.adsabs.harvard.edu/abs/2001AJ....122..288H} {122, 288}

\bibitem[\protect\citeauthoryear{{Ilbert} et~al.,}{{Ilbert}
  et~al.}{2006}]{Ilbert_et_al_2006}
{Ilbert} O.,  et~al., 2006, \mn@doi [\aap] {10.1051/0004-6361:20065138}, \href
  {http://adsabs.harvard.edu/abs/2006A%26A...457..841I} {457, 841}

\bibitem[\protect\citeauthoryear{{Jaff{\'e}}, {Smith}, {Candlish}, {Poggianti},
  {Sheen}  \& {Verheijen}}{{Jaff{\'e}} et~al.}{2015}]{Jaffe2015}
{Jaff{\'e}} Y.~L.,  {Smith} R.,  {Candlish} G.~N.,  {Poggianti} B.~M.,  {Sheen}
  Y.-K.,   {Verheijen} M.~A.~W.,  2015, \mn@doi [\mnras]
  {10.1093/mnras/stv100}, \href
  {http://adsabs.harvard.edu/abs/2015MNRAS.448.1715J} {448, 1715}

\bibitem[\protect\citeauthoryear{{Jones} et~al.,}{{Jones}
  et~al.}{2004}]{Jones2004MNRAS}
{Jones} D.~H.,  et~al., 2004, \mn@doi [\mnras]
  {10.1111/j.1365-2966.2004.08353.x}, 355, 747

\bibitem[\protect\citeauthoryear{{Kartaltepe} et~al.,}{{Kartaltepe}
  et~al.}{2015}]{Kartaltepe2015}
{Kartaltepe} J.~S.,  et~al., 2015, \mn@doi [\apjs]
  {10.1088/0067-0049/221/1/11}, \href
  {https://ui.adsabs.harvard.edu/abs/2015ApJS..221...11K} {221, 11}

\bibitem[\protect\citeauthoryear{{Kelvin} et~al.,}{{Kelvin}
  et~al.}{2012}]{Kelvin2012}
{Kelvin} L.~S.,  et~al., 2012, \mn@doi [\mnras]
  {10.1111/j.1365-2966.2012.20355.x}, \href
  {https://ui.adsabs.harvard.edu/abs/2012MNRAS.421.1007K} {421, 1007}

\bibitem[\protect\citeauthoryear{{Kennedy} et~al.,}{{Kennedy}
  et~al.}{2015}]{Kennedy2015}
{Kennedy} R.,  et~al., 2015, \mn@doi [\mnras] {10.1093/mnras/stv2032}, \href
  {http://adsabs.harvard.edu/abs/2015MNRAS.454..806K} {454, 806}

\bibitem[\protect\citeauthoryear{{Kennicutt}}{{Kennicutt}}{1992}]{Kennicutt1992ApJ}
{Kennicutt} Robert~C. J.,  1992, \mn@doi [\apj] {10.1086/171154}, \href
  {https://ui.adsabs.harvard.edu/abs/1992ApJ...388..310K} {388, 310}

\bibitem[\protect\citeauthoryear{{Kennicutt}}{{Kennicutt}}{1998}]{Kennicutt1998}
{Kennicutt} Jr. R.~C.,  1998, \mn@doi [\apj] {10.1086/305588}, \href
  {http://adsabs.harvard.edu/abs/1998ApJ...498..541K} {498, 541}

\bibitem[\protect\citeauthoryear{{Khostovan} et~al.,}{{Khostovan}
  et~al.}{2020}]{Khostovan2020MNRAS}
{Khostovan} A.~A.,  et~al., 2020, \mn@doi [\mnras] {10.1093/mnras/staa175},
  \href {https://ui.adsabs.harvard.edu/abs/2020MNRAS.493.3966K} {493, 3966}

\bibitem[\protect\citeauthoryear{{Koyama} et~al.,}{{Koyama}
  et~al.}{2013}]{Koyama2013}
{Koyama} Y.,  et~al., 2013, \mn@doi [\mnras] {10.1093/mnras/stt1035}, \href
  {https://ui.adsabs.harvard.edu/abs/2013MNRAS.434..423K} {434, 423}

\bibitem[\protect\citeauthoryear{{La Barbera}, {de Carvalho}, {de La Rosa},
  {Lopes}, {Kohl-Moreira}  \& {Capelato}}{{La Barbera}
  et~al.}{2010}]{LaBarbera2010}
{La Barbera} F.,  {de Carvalho} R.~R.,  {de La Rosa} I.~G.,  {Lopes} P.~A.~A.,
  {Kohl-Moreira} J.~L.,   {Capelato} H.~V.,  2010, \mn@doi [\mnras]
  {10.1111/j.1365-2966.2010.16850.x}, \href
  {https://ui.adsabs.harvard.edu/abs/2010MNRAS.408.1313L} {408, 1313}

\bibitem[\protect\citeauthoryear{{Lagan{\'a}} \& {Ulmer}}{{Lagan{\'a}} \&
  {Ulmer}}{2018}]{Lagana2018}
{Lagan{\'a}} T.~F.,  {Ulmer} M.~P.,  2018, \mn@doi [\mnras]
  {10.1093/mnras/stx3210}, \href
  {https://ui.adsabs.harvard.edu/abs/2018MNRAS.475..523L} {475, 523}

\bibitem[\protect\citeauthoryear{{Larson}}{{Larson}}{1974}]{Larson74}
{Larson} R.~B.,  1974, \mn@doi [\mnras] {10.1093/mnras/169.2.229}, \href
  {https://ui.adsabs.harvard.edu/abs/1974MNRAS.169..229L} {169, 229}

\bibitem[\protect\citeauthoryear{{Lauer} et~al.,}{{Lauer}
  et~al.}{1995}]{Lauer1995}
{Lauer} T.~R.,  et~al., 1995, \mn@doi [\aj] {10.1086/117719}, \href
  {https://ui.adsabs.harvard.edu/abs/1995AJ....110.2622L} {110, 2622}

\bibitem[\protect\citeauthoryear{{Lee}, {Lee}, {Kim}, {Hwang}, {Park}  \&
  {Choi}}{{Lee} et~al.}{2007}]{Lee2007}
{Lee} J.~H.,  {Lee} M.~G.,  {Kim} T.,  {Hwang} H.~S.,  {Park} C.,   {Choi}
  Y.-Y.,  2007, \mn@doi [\apjl] {10.1086/518887}, \href
  {http://adsabs.harvard.edu/abs/2007ApJ...663L..69L} {663, L69}

\bibitem[\protect\citeauthoryear{{Leonard} \& {King}}{{Leonard} \&
  {King}}{2010}]{LeonardKing2010}
{Leonard} A.,  {King} L.~J.,  2010, \mn@doi [Monthly Notices of the Royal
  Astronomical Society] {10.1111/j.1365-2966.2010.16573.x}, \href
  {https://ui.adsabs.harvard.edu/abs/2010MNRAS.405.1854L} {405, 1854}

\bibitem[\protect\citeauthoryear{{Lintott} et~al.,}{{Lintott}
  et~al.}{2008}]{Lintott2008}
{Lintott} C.~J.,  et~al., 2008, \mn@doi [\mnras]
  {10.1111/j.1365-2966.2008.13689.x}, \href
  {https://ui.adsabs.harvard.edu/abs/2008MNRAS.389.1179L} {389, 1179}

\bibitem[\protect\citeauthoryear{{Liu} et~al.,}{{Liu} et~al.}{2011}]{liu2011}
{Liu} S.-F.,  et~al., 2011, \mn@doi [\aj] {10.1088/0004-6256/141/3/99}, \href
  {https://ui.adsabs.harvard.edu/abs/2011AJ....141...99L} {141, 99}

\bibitem[\protect\citeauthoryear{{Liu}, {Hao}, {Wang}  \& {Yang}}{{Liu}
  et~al.}{2019}]{Liu2019}
{Liu} C.,  {Hao} L.,  {Wang} H.,   {Yang} X.,  2019, \mn@doi [\apj]
  {10.3847/1538-4357/ab1ea0}, \href
  {https://ui.adsabs.harvard.edu/abs/2019ApJ...878...69L} {878, 69}

\bibitem[\protect\citeauthoryear{{{\L}okas}, {Wojtak}, {Gottl{\"o}ber}, {Mamon}
   \& {Prada}}{{{\L}okas} et~al.}{2006}]{Lokas2006}
{{\L}okas} E.~L.,  {Wojtak} R.,  {Gottl{\"o}ber} S.,  {Mamon} G.~A.,   {Prada}
  F.,  2006, \mn@doi [\mnras] {10.1111/j.1365-2966.2006.10151.x}, \href
  {http://adsabs.harvard.edu/abs/2006MNRAS.367.1463L} {367, 1463}

\bibitem[\protect\citeauthoryear{{Lotz}, {Primack}  \& {Madau}}{{Lotz}
  et~al.}{2004}]{Lotz2004AJ....128..163L}
{Lotz} J.~M.,  {Primack} J.,   {Madau} P.,  2004, \mn@doi [\aj]
  {10.1086/421849}, \href
  {https://ui.adsabs.harvard.edu/abs/2004AJ....128..163L} {128, 163}

\bibitem[\protect\citeauthoryear{{Lotz} et~al.,}{{Lotz}
  et~al.}{2008}]{Lotz2008}
{Lotz} J.~M.,  et~al., 2008, \mn@doi [\apj] {10.1086/523659}, \href
  {http://adsabs.harvard.edu/abs/2008ApJ...672..177L} {672, 177}

\bibitem[\protect\citeauthoryear{{Ly} et~al.,}{{Ly} et~al.}{2007}]{Ly2007ApJ}
{Ly} C.,  et~al., 2007, \mn@doi [\apj] {10.1086/510828}, \href
  {https://ui.adsabs.harvard.edu/abs/2007ApJ...657..738L} {657, 738}

\bibitem[\protect\citeauthoryear{{Ly}, {Malkan}, {Kashikawa}, {Ota},
  {Shimasaku}, {Iye}  \& {Currie}}{{Ly} et~al.}{2012}]{Ly2012ApJ}
{Ly} C.,  {Malkan} M.~A.,  {Kashikawa} N.,  {Ota} K.,  {Shimasaku} K.,  {Iye}
  M.,   {Currie} T.,  2012, \mn@doi [\apjl] {10.1088/2041-8205/747/1/L16},
  \href {https://ui.adsabs.harvard.edu/abs/2012ApJ...747L..16L} {747, L16}

\bibitem[\protect\citeauthoryear{{Mendes de Oliveira} et~al.,}{{Mendes de
  Oliveira} et~al.}{2019}]{MendesdeOliveira2019}
{Mendes de Oliveira} C.,  et~al., 2019, \mn@doi [\mnras]
  {10.1093/mnras/stz1985}, \href
  {https://ui.adsabs.harvard.edu/abs/2019MNRAS.tmp.2048M} {p.~2048}

\bibitem[\protect\citeauthoryear{{Misgeld} \& {Hilker}}{{Misgeld} \&
  {Hilker}}{2011}]{MisgeldHilker2011}
{Misgeld} I.,  {Hilker} M.,  2011, \mn@doi [\mnras]
  {10.1111/j.1365-2966.2011.18669.x}, \href
  {http://adsabs.harvard.edu/abs/2011MNRAS.414.3699M} {414, 3699}

\bibitem[\protect\citeauthoryear{{Misgeld}, {Mieske}, {Hilker}, {Richtler},
  {Georgiev}  \& {Schuberth}}{{Misgeld} et~al.}{2011}]{Misgeld2011}
{Misgeld} I.,  {Mieske} S.,  {Hilker} M.,  {Richtler} T.,  {Georgiev} I.~Y.,
  {Schuberth} Y.,  2011, \mn@doi [\aap] {10.1051/0004-6361/201116728}, \href
  {http://adsabs.harvard.edu/abs/2011A%26A...531A...4M} {531, A4}

\bibitem[\protect\citeauthoryear{{Moffat}}{{Moffat}}{1969}]{Moffat1969A&A}
{Moffat} A.~F.~J.,  1969, \aap, \href
  {https://ui.adsabs.harvard.edu/abs/1969A&A.....3..455M} {3, 455}

\bibitem[\protect\citeauthoryear{{Molino} et~al.,}{{Molino}
  et~al.}{2019}]{Molino2019}
{Molino} A.,  et~al., 2019, arXiv e-prints, \href
  {https://ui.adsabs.harvard.edu/abs/2019arXiv190706315M} {p. arXiv:1907.06315}

\bibitem[\protect\citeauthoryear{{Moore}, {Lake}  \& {Katz}}{{Moore}
  et~al.}{1998}]{Moore98}
{Moore} B.,  {Lake} G.,   {Katz} N.,  1998, \mn@doi [\apj] {10.1086/305264},
  \href {https://ui.adsabs.harvard.edu/abs/1998ApJ...495..139M} {495, 139}

\bibitem[\protect\citeauthoryear{{Moore}, {Lake}, {Quinn}  \& {Stadel}}{{Moore}
  et~al.}{1999}]{Moore99}
{Moore} B.,  {Lake} G.,  {Quinn} T.,   {Stadel} J.,  1999, \mn@doi [\mnras]
  {10.1046/j.1365-8711.1999.02345.x}, \href
  {https://ui.adsabs.harvard.edu/abs/1999MNRAS.304..465M} {304, 465}

\bibitem[\protect\citeauthoryear{{Munari}, {Biviano}, {Borgani}, {Murante}  \&
  {Fabjan}}{{Munari} et~al.}{2013}]{Munari2013}
{Munari} E.,  {Biviano} A.,  {Borgani} S.,  {Murante} G.,   {Fabjan} D.,  2013,
  \mn@doi [\mnras] {10.1093/mnras/stt049}, \href
  {https://ui.adsabs.harvard.edu/abs/2013MNRAS.430.2638M} {430, 2638}

\bibitem[\protect\citeauthoryear{{Navarro}, {Frenk}  \& {White}}{{Navarro}
  et~al.}{1996}]{NFW1996}
{Navarro} J.~F.,  {Frenk} C.~S.,   {White} S.~D.~M.,  1996, \mn@doi [\apj]
  {10.1086/177173}, \href {http://adsabs.harvard.edu/abs/1996ApJ...462..563N}
  {462, 563}

\bibitem[\protect\citeauthoryear{{Nelson} et~al.,}{{Nelson}
  et~al.}{2018}]{Nelson2018}
{Nelson} D.,  et~al., 2018, \mn@doi [\mnras] {10.1093/mnras/stx3040}, \href
  {https://ui.adsabs.harvard.edu/abs/2018MNRAS.475..624N} {475, 624}

\bibitem[\protect\citeauthoryear{{Oemler}}{{Oemler}}{1974}]{Oemler74}
{Oemler} Augustus J.,  1974, \mn@doi [\apj] {10.1086/153216}, \href
  {https://ui.adsabs.harvard.edu/abs/1974ApJ...194....1O} {194, 1}

\bibitem[\protect\citeauthoryear{{Olave-Rojas}, {Cerulo}, {Demarco},
  {Jaff{\'e}}, {Mercurio}, {Rosati}, {Balestra}  \& {Nonino}}{{Olave-Rojas}
  et~al.}{2018}]{Olave-Rojas2018}
{Olave-Rojas} D.,  {Cerulo} P.,  {Demarco} R.,  {Jaff{\'e}} Y.~L.,  {Mercurio}
  A.,  {Rosati} P.,  {Balestra} I.,   {Nonino} M.,  2018, \mn@doi [\mnras]
  {10.1093/mnras/sty1669}, \href
  {https://ui.adsabs.harvard.edu/abs/2018MNRAS.479.2328O} {479, 2328}

\bibitem[\protect\citeauthoryear{{Owers} et~al.,}{{Owers}
  et~al.}{2019}]{Owers2019ApJ}
{Owers} M.~S.,  et~al., 2019, \mn@doi [\apj] {10.3847/1538-4357/ab0201}, \href
  {https://ui.adsabs.harvard.edu/abs/2019ApJ...873...52O} {873, 52}

\bibitem[\protect\citeauthoryear{{Pallero}, {G{\'o}mez}, {Padilla},
  {Torres-Flores}, {Demarco}, {Cerulo}  \& {Olave-Rojas}}{{Pallero}
  et~al.}{2019}]{Pallero2019}
{Pallero} D.,  {G{\'o}mez} F.~A.,  {Padilla} N.~D.,  {Torres-Flores} S.,
  {Demarco} R.,  {Cerulo} P.,   {Olave-Rojas} D.,  2019, \mn@doi [\mnras]
  {10.1093/mnras/stz1745}, \href
  {https://ui.adsabs.harvard.edu/abs/2019MNRAS.488..847P} {488, 847}

\bibitem[\protect\citeauthoryear{{Papovich}, {Dickinson}, {Giavalisco},
  {Conselice}  \& {Ferguson}}{{Papovich} et~al.}{2005}]{Papovich2005ApJ}
{Papovich} C.,  {Dickinson} M.,  {Giavalisco} M.,  {Conselice} C.~J.,
  {Ferguson} H.~C.,  2005, \mn@doi [\apj] {10.1086/429120}, \href
  {https://ui.adsabs.harvard.edu/abs/2005ApJ...631..101P} {631, 101}

\bibitem[\protect\citeauthoryear{{Park} \& {Choi}}{{Park} \&
  {Choi}}{2005}]{Park2005}
{Park} C.,  {Choi} Y.-Y.,  2005, \mn@doi [\apjl] {10.1086/499243}, \href
  {http://adsabs.harvard.edu/abs/2005ApJ...635L..29P} {635, L29}

\bibitem[\protect\citeauthoryear{{Pascual}, {Gallego}  \& {Zamorano}}{{Pascual}
  et~al.}{2007}]{Pascual2007}
{Pascual} S.,  {Gallego} J.,   {Zamorano} J.,  2007, \mn@doi [\pasp]
  {10.1086/510600}, \href {http://adsabs.harvard.edu/abs/2007PASP..119...30P}
  {119, 30}

\bibitem[\protect\citeauthoryear{{Paulino-Afonso} et~al.,}{{Paulino-Afonso}
  et~al.}{2019}]{Paulono-Afonso2019}
{Paulino-Afonso} A.,  et~al., 2019, \mn@doi [\aap]
  {10.1051/0004-6361/201935137}, \href
  {https://ui.adsabs.harvard.edu/abs/2019A&A...630A..57P} {630, A57}

\bibitem[\protect\citeauthoryear{{Peng}, {Ho}, {Impey}  \& {Rix}}{{Peng}
  et~al.}{2002}]{Peng2002}
{Peng} C.~Y.,  {Ho} L.~C.,  {Impey} C.~D.,   {Rix} H.-W.,  2002, \mn@doi [\aj]
  {10.1086/340952}, \href {http://adsabs.harvard.edu/abs/2002AJ....124..266P}
  {124, 266}

\bibitem[\protect\citeauthoryear{{Peng}, {Ho}, {Impey}  \& {Rix}}{{Peng}
  et~al.}{2010a}]{Peng2010}
{Peng} C.~Y.,  {Ho} L.~C.,  {Impey} C.~D.,   {Rix} H.-W.,  2010a, \mn@doi [\aj]
  {10.1088/0004-6256/139/6/2097}, \href
  {https://ui.adsabs.harvard.edu/abs/2010AJ....139.2097P} {139, 2097}

\bibitem[\protect\citeauthoryear{{Peng} et~al.,}{{Peng}
  et~al.}{2010b}]{Peng2010ApJ}
{Peng} Y.-j.,  et~al., 2010b, \mn@doi [\apj] {10.1088/0004-637X/721/1/193},
  \href {https://ui.adsabs.harvard.edu/abs/2010ApJ...721..193P} {721, 193}

\bibitem[\protect\citeauthoryear{{Peng}, {Maiolino}  \& {Cochrane}}{{Peng}
  et~al.}{2015}]{Peng2015}
{Peng} Y.,  {Maiolino} R.,   {Cochrane} R.,  2015, \mn@doi [\nat]
  {10.1038/nature14439}, \href
  {https://ui.adsabs.harvard.edu/abs/2015Natur.521..192P} {521, 192}

\bibitem[\protect\citeauthoryear{{Pinkney}, {Roettiger}, {Burns}  \&
  {Bird}}{{Pinkney} et~al.}{1996}]{Pinkney1996}
{Pinkney} J.,  {Roettiger} K.,  {Burns} J.~O.,   {Bird} C.~M.,  1996, \mn@doi
  [\apjs] {10.1086/192290}, \href
  {http://adsabs.harvard.edu/abs/1996ApJS..104....1P} {104, 1}

\bibitem[\protect\citeauthoryear{{Poggianti} et~al.,}{{Poggianti}
  et~al.}{2001}]{Poggianti01}
{Poggianti} B.~M.,  et~al., 2001, \mn@doi [\apj] {10.1086/323217}, \href
  {https://ui.adsabs.harvard.edu/abs/2001ApJ...562..689P} {562, 689}

\bibitem[\protect\citeauthoryear{{Postman} \& {Geller}}{{Postman} \&
  {Geller}}{1984}]{Postman84}
{Postman} M.,  {Geller} M.~J.,  1984, \mn@doi [\apj] {10.1086/162078}, \href
  {https://ui.adsabs.harvard.edu/abs/1984ApJ...281...95P} {281, 95}

\bibitem[\protect\citeauthoryear{{Postman} et~al.,}{{Postman}
  et~al.}{2005}]{Postman05}
{Postman} M.,  et~al., 2005, \mn@doi [\apj] {10.1086/428881}, \href
  {https://ui.adsabs.harvard.edu/abs/2005ApJ...623..721P} {623, 721}

\bibitem[\protect\citeauthoryear{{Psychogyios} et~al.,}{{Psychogyios}
  et~al.}{2019}]{Psychogyios2019}
{Psychogyios} A.,  et~al., 2019, arXiv e-prints, \href
  {https://ui.adsabs.harvard.edu/abs/2019arXiv190903256P} {p. arXiv:1909.03256}

\bibitem[\protect\citeauthoryear{{Quilis}, {Moore}  \& {Bower}}{{Quilis}
  et~al.}{2000}]{Quilis00}
{Quilis} V.,  {Moore} B.,   {Bower} R.,  2000, \mn@doi [Science]
  {10.1126/science.288.5471.1617}, \href
  {https://ui.adsabs.harvard.edu/abs/2000Sci...288.1617Q} {288, 1617}

\bibitem[\protect\citeauthoryear{{Richter}}{{Richter}}{1987}]{Richter1987A&AS}
{Richter} O.~G.,  1987, \aaps, \href
  {https://ui.adsabs.harvard.edu/abs/1987A&AS...67..237R} {67, 237}

\bibitem[\protect\citeauthoryear{{Rodr{\'\i}guez-Mu{\~n}oz}
  et~al.,}{{Rodr{\'\i}guez-Mu{\~n}oz} et~al.}{2019}]{Rodrigo_Munoz2019}
{Rodr{\'\i}guez-Mu{\~n}oz} L.,  et~al., 2019, \mn@doi [\mnras]
  {10.1093/mnras/sty3335}, \href
  {https://ui.adsabs.harvard.edu/abs/2019MNRAS.485..586R} {485, 586}

\bibitem[\protect\citeauthoryear{{Ruel} et~al.,}{{Ruel}
  et~al.}{2014}]{Ruel2014}
{Ruel} J.,  et~al., 2014, \mn@doi [\apj] {10.1088/0004-637X/792/1/45}, \href
  {https://ui.adsabs.harvard.edu/abs/2014ApJ...792...45R} {792, 45}

\bibitem[\protect\citeauthoryear{{Salim}}{{Salim}}{2014}]{Salin2014}
{Salim} S.,  2014, \mn@doi [Serbian Astronomical Journal]
  {10.2298/SAJ1489001S}, \href
  {https://ui.adsabs.harvard.edu/abs/2014SerAJ.189....1S} {189, 1}

\bibitem[\protect\citeauthoryear{{Salpeter}}{{Salpeter}}{1955}]{Salpeter1955}
{Salpeter} E.~E.,  1955, \mn@doi [\apj] {10.1086/145971}, \href
  {http://adsabs.harvard.edu/abs/1955ApJ...121..161S} {121, 161}

\bibitem[\protect\citeauthoryear{{Schawinski} et~al.,}{{Schawinski}
  et~al.}{2014}]{Schawinski2014MNRAS}
{Schawinski} K.,  et~al., 2014, \mn@doi [\mnras] {10.1093/mnras/stu327}, \href
  {https://ui.adsabs.harvard.edu/abs/2014MNRAS.440..889S} {440, 889}

\bibitem[\protect\citeauthoryear{{Schlegel}, {Finkbeiner}  \&
  {Davis}}{{Schlegel} et~al.}{1998}]{Schlegel1998ApJ}
{Schlegel} D.~J.,  {Finkbeiner} D.~P.,   {Davis} M.,  1998, \mn@doi [\apj]
  {10.1086/305772}, \href
  {https://ui.adsabs.harvard.edu/abs/1998ApJ...500..525S} {500, 525}

\bibitem[\protect\citeauthoryear{{S{\'e}rsic}}{{S{\'e}rsic}}{1963}]{Sersic1963}
{S{\'e}rsic} J.~L.,  1963, Boletin de la Asociacion Argentina de Astronomia La
  Plata Argentina, \href
  {https://ui.adsabs.harvard.edu/abs/1963BAAA....6...41S} {6, 41}

\bibitem[\protect\citeauthoryear{{Simmons} et~al.,}{{Simmons}
  et~al.}{2017}]{Simmons2017}
{Simmons} B.~D.,  et~al., 2017, \mn@doi [\mnras] {10.1093/mnras/stw2587}, \href
  {https://ui.adsabs.harvard.edu/abs/2017MNRAS.464.4420S} {464, 4420}

\bibitem[\protect\citeauthoryear{{Skrutskie} et~al.,}{{Skrutskie}
  et~al.}{1997}]{Skrutskie1997ASSL}
{Skrutskie} M.~F.,  et~al., 1997, {The Two Micron All Sky Survey (2MASS):
  Overview and Status.}.
p.~25, \mn@doi{10.1007/978-94-011-5784-1_4}

\bibitem[\protect\citeauthoryear{{Smith} et~al.,}{{Smith}
  et~al.}{2015}]{Smith15}
{Smith} R.,  et~al., 2015, \mn@doi [\mnras] {10.1093/mnras/stv2082}, \href
  {https://ui.adsabs.harvard.edu/abs/2015MNRAS.454.2502S} {454, 2502}

\bibitem[\protect\citeauthoryear{{Sobral}, {Best}, {Matsuda}, {Smail}, {Geach}
  \& {Cirasuolo}}{{Sobral} et~al.}{2012}]{Sobral2012MNRAS}
{Sobral} D.,  {Best} P.~N.,  {Matsuda} Y.,  {Smail} I.,  {Geach} J.~E.,
  {Cirasuolo} M.,  2012, \mn@doi [\mnras] {10.1111/j.1365-2966.2011.19977.x},
  \href {https://ui.adsabs.harvard.edu/abs/2012MNRAS.420.1926S} {420, 1926}

\bibitem[\protect\citeauthoryear{{Sohn}, {Hwang}, {Geller}, {Diaferio},
  {Rines}, {Lee}  \& {Lee}}{{Sohn} et~al.}{2015}]{Sohn2015}
{Sohn} J.,  {Hwang} H.~S.,  {Geller} M.~J.,  {Diaferio} A.,  {Rines} K.~J.,
  {Lee} M.~G.,   {Lee} G.-H.,  2015, \mn@doi [Journal of Korean Astronomical
  Society] {10.5303/JKAS.2015.48.6.381}, \href
  {https://ui.adsabs.harvard.edu/abs/2015JKAS...48..381S} {48, 381}

\bibitem[\protect\citeauthoryear{{Spergel} et~al.,}{{Spergel}
  et~al.}{2003}]{Spergel2003ApJS}
{Spergel} D.~N.,  et~al., 2003, \mn@doi [\apjs] {10.1086/377226}, \href
  {https://ui.adsabs.harvard.edu/abs/2003ApJS..148..175S} {148, 175}

\bibitem[\protect\citeauthoryear{{Stein}}{{Stein}}{1996}]{Stein1996A&AS}
{Stein} P.,  1996, \aaps, \href
  {https://ui.adsabs.harvard.edu/abs/1996A&AS..116..203S} {116, 203}

\bibitem[\protect\citeauthoryear{{Stein}}{{Stein}}{1997}]{Stein1997}
{Stein} P.,  1997, \aap, \href
  {https://ui.adsabs.harvard.edu/abs/1997A&A...317..670S} {317, 670}

\bibitem[\protect\citeauthoryear{{Taylor} et~al.,}{{Taylor}
  et~al.}{2011}]{Taylor2011}
{Taylor} E.~N.,  et~al., 2011, \mn@doi [\mnras]
  {10.1111/j.1365-2966.2011.19536.x}, \href
  {http://adsabs.harvard.edu/abs/2011MNRAS.418.1587T} {418, 1587}

\bibitem[\protect\citeauthoryear{{Ventimiglia}, {Arnaboldi}  \&
  {Gerhard}}{{Ventimiglia} et~al.}{2011}]{Ventimiglia2011}
{Ventimiglia} G.,  {Arnaboldi} M.,   {Gerhard} O.,  2011, \mn@doi [\aap]
  {10.1051/0004-6361/201015982}, \href
  {http://adsabs.harvard.edu/abs/2011A%26A...528A..24V} {528, A24}

\bibitem[\protect\citeauthoryear{{Vika}, {Bamford}, {H{\"a}u{\ss}ler}, {Rojas},
  {Borch}  \& {Nichol}}{{Vika} et~al.}{2013}]{Vika2013}
{Vika} M.,  {Bamford} S.~P.,  {H{\"a}u{\ss}ler} B.,  {Rojas} A.~L.,  {Borch}
  A.,   {Nichol} R.~C.,  2013, \mn@doi [\mnras] {10.1093/mnras/stt1320}, \href
  {http://adsabs.harvard.edu/abs/2013MNRAS.435..623V} {435, 623}

\bibitem[\protect\citeauthoryear{{Vika}, {Bamford}, {H{\"a}u{\ss}ler}  \&
  {Rojas}}{{Vika} et~al.}{2014}]{Vika2014MNRAS}
{Vika} M.,  {Bamford} S.~P.,  {H{\"a}u{\ss}ler} B.,   {Rojas} A.~L.,  2014,
  \mn@doi [\mnras] {10.1093/mnras/stu1696}, \href
  {https://ui.adsabs.harvard.edu/abs/2014MNRAS.444.3603V} {444, 3603}

\bibitem[\protect\citeauthoryear{{Vika}, {Vulcani}, {Bamford},
  {H{\"a}u{\ss}ler}  \& {Rojas}}{{Vika} et~al.}{2015}]{Vika2015}
{Vika} M.,  {Vulcani} B.,  {Bamford} S.~P.,  {H{\"a}u{\ss}ler} B.,   {Rojas}
  A.~L.,  2015, \mn@doi [\aap] {10.1051/0004-6361/201425174}, \href
  {http://adsabs.harvard.edu/abs/2015A%26A...577A..97V} {577, A97}

\bibitem[\protect\citeauthoryear{{Vilella-Rojo} et~al.,}{{Vilella-Rojo}
  et~al.}{2015}]{Vilella2015}
{Vilella-Rojo} G.,  et~al., 2015, \mn@doi [\aap] {10.1051/0004-6361/201526374},
  \href {http://adsabs.harvard.edu/abs/2015A%26A...580A..47V} {580, A47}

\bibitem[\protect\citeauthoryear{{Vollmer}, {Cayatte}, {Balkowski}  \&
  {Duschl}}{{Vollmer} et~al.}{2001}]{Vollmer01}
{Vollmer} B.,  {Cayatte} V.,  {Balkowski} C.,   {Duschl} W.~J.,  2001, \mn@doi
  [\apj] {10.1086/323368}, \href
  {https://ui.adsabs.harvard.edu/abs/2001ApJ...561..708V} {561, 708}

\bibitem[\protect\citeauthoryear{{Vulcani} et~al.,}{{Vulcani}
  et~al.}{2011}]{Vulcani2011MNRAS}
{Vulcani} B.,  et~al., 2011, \mn@doi [\mnras]
  {10.1111/j.1365-2966.2010.17904.x}, \href
  {https://ui.adsabs.harvard.edu/abs/2011MNRAS.412..246V} {412, 246}

\bibitem[\protect\citeauthoryear{{Vulcani} et~al.,}{{Vulcani}
  et~al.}{2014}]{Vulcani2014}
{Vulcani} B.,  et~al., 2014, \mn@doi [\apj] {10.1088/0004-637X/797/1/62}, \href
  {http://adsabs.harvard.edu/abs/2014ApJ...797...62V} {797, 62}

\bibitem[\protect\citeauthoryear{{Weinmann}, {Kauffmann}, {von der Linden}  \&
  {De Lucia}}{{Weinmann} et~al.}{2010}]{Weinmann2010}
{Weinmann} S.~M.,  {Kauffmann} G.,  {von der Linden} A.,   {De Lucia} G.,
  2010, \mn@doi [\mnras] {10.1111/j.1365-2966.2010.16855.x}, \href
  {https://ui.adsabs.harvard.edu/abs/2010MNRAS.406.2249W} {406, 2249}

\bibitem[\protect\citeauthoryear{{Wetzel}, {Tinker}  \& {Conroy}}{{Wetzel}
  et~al.}{2012}]{wetzel2012}
{Wetzel} A.~R.,  {Tinker} J.~L.,   {Conroy} C.,  2012, \mn@doi [\mnras]
  {10.1111/j.1365-2966.2012.21188.x}, \href
  {http://adsabs.harvard.edu/abs/2012MNRAS.424..232W} {424, 232}

\bibitem[\protect\citeauthoryear{{Wetzel}, {Tinker}, {Conroy}  \& {van den
  Bosch}}{{Wetzel} et~al.}{2013}]{Wetzel2013}
{Wetzel} A.~R.,  {Tinker} J.~L.,  {Conroy} C.,   {van den Bosch} F.~C.,  2013,
  \mn@doi [\mnras] {10.1093/mnras/stt469}, \href
  {http://adsabs.harvard.edu/abs/2013MNRAS.432..336W} {432, 336}

\bibitem[\protect\citeauthoryear{{Wetzel}, {Tinker}, {Conroy}  \& {van den
  Bosch}}{{Wetzel} et~al.}{2014}]{Wetzel2014}
{Wetzel} A.~R.,  {Tinker} J.~L.,  {Conroy} C.,   {van den Bosch} F.~C.,  2014,
  \mn@doi [\mnras] {10.1093/mnras/stu122}, \href
  {http://adsabs.harvard.edu/abs/2014MNRAS.439.2687W} {439, 2687}

\bibitem[\protect\citeauthoryear{{White} et~al.,}{{White}
  et~al.}{2015}]{White2015}
{White} J.~A.,  et~al., 2015, \mn@doi [\mnras] {10.1093/mnras/stv1831}, \href
  {https://ui.adsabs.harvard.edu/abs/2015MNRAS.453.2718W} {453, 2718}

\bibitem[\protect\citeauthoryear{{Whitmore} \& {Gilmore}}{{Whitmore} \&
  {Gilmore}}{1991}]{Whitmore1991ApJ}
{Whitmore} B.~C.,  {Gilmore} D.~M.,  1991, \mn@doi [\apj] {10.1086/169602},
  \href {https://ui.adsabs.harvard.edu/abs/1991ApJ...367...64W} {367, 64}

\bibitem[\protect\citeauthoryear{{Wright} et~al.,}{{Wright}
  et~al.}{2010}]{Wright2010}
{Wright} E.~L.,  et~al., 2010, \mn@doi [\aj] {10.1088/0004-6256/140/6/1868},
  \href {http://adsabs.harvard.edu/abs/2010AJ....140.1868W} {140, 1868}

\bibitem[\protect\citeauthoryear{{Wright}, {Lagos}, {Davies}, {Power},
  {Trayford}  \& {Wong}}{{Wright} et~al.}{2019}]{Wright2019}
{Wright} R.~J.,  {Lagos} C. d.~P.,  {Davies} L. J.~M.,  {Power} C.,  {Trayford}
  J.~W.,   {Wong} O.~I.,  2019, \mn@doi [\mnras] {10.1093/mnras/stz1410}, \href
  {https://ui.adsabs.harvard.edu/abs/2019MNRAS.487.3740W} {487, 3740}

\bibitem[\protect\citeauthoryear{{Yan}, {Yuan}, {Zhang}  \& {Zhou}}{{Yan}
  et~al.}{2014}]{Yan2014}
{Yan} P.-F.,  {Yuan} Q.-R.,  {Zhang} L.,   {Zhou} X.,  2014, \mn@doi [\aj]
  {10.1088/0004-6256/147/5/106}, \href
  {https://ui.adsabs.harvard.edu/abs/2014AJ....147..106Y} {147, 106}

\bibitem[\protect\citeauthoryear{{da Cunha}, {Charlot}, {Dunne}, {Smith}  \&
  {Rowlands}}{{da Cunha} et~al.}{2012}]{daCunha2012}
{da Cunha} E.,  {Charlot} S.,  {Dunne} L.,  {Smith} D.,   {Rowlands} K.,  2012,
  in {Tuffs} R.~J.,  {Popescu} C.~C.,  eds,  IAU Symposium Vol. 284, The
  Spectral Energy Distribution of Galaxies - SED 2011. pp 292--296 (\mn@eprint
  {arXiv} {1111.3961}), \mn@doi{10.1017/S1743921312009283}

\bibitem[\protect\citeauthoryear{{van der Burg} et~al.,}{{van der Burg}
  et~al.}{2020}]{vanderBurg2020A&A}
{van der Burg} R. F.~J.,  et~al., 2020, \mn@doi [\aap]
  {10.1051/0004-6361/202037754}, \href
  {https://ui.adsabs.harvard.edu/abs/2020A&A...638A.112V} {638, A112}

\makeatother
\end{thebibliography}


%%%%%%%%%%%%%%%%%%%%%%%%%%%%%%%%%%%%%%%%%%%%%%%%%%

%%%%%%%%%%%%%%%%% APPENDICES %%%%%%%%%%%%%%%%%%%%%

\appendix 

\section{Data Tables}
\label{appendix:data}

In this Appendix we present an example of the tables with the photometric information of each galaxy as well as their parameters determined in this work. Details about how these parameters were determined are found in Sections \ref{sec:morphology}, \ref{sec:mass} and \ref{sec:sfr}. Full versions of these tables can be downloaded from the online version of the paper.

%\begin{ThreePartTable}
\begin{table*} 
\caption{Magnitudes estimated in this work using MegaMorph project. Column 1 is the ID of each galaxy, and the columns 2 to 13 are the magnitudes in the 12 S-PLUS filters, without the correction for Galactic extinction. A full version of this table, for all galaxies, is available as supporting information in the online version of this manuscript.} \label{tab:Mag_MegaMorph} 
\resizebox{\textwidth}{!}{
\begin{tabular}{lllllllllllllllll}
\hline
\hline
ID & $u$ & J0378 & J0395 & J0410 & J0430 & $g$ & J0515 & $r$ & J0660 & $i$ & J0861 & $z$ \\
\hline
1  & 15.41$\pm$0.01 & 15.05$\pm$0.02 & 14.81$\pm$0.02 & 14.62$\pm$0.01 & 14.43$\pm$0.01 & 14.12$\pm$0.01 & 13.91$\pm$0.1 & 13.45$\pm$0.1 & 13.19$\pm$0.01 & 13.09$\pm$0.01 & 12.93$\pm$0.02 & 12.87$\pm$0.01\\
\hline

\end{tabular}
}
\end{table*}

\begin{table*}
\caption{S\'ersic index estimated in this work using MegaMorph project. Column 1 is the ID of each galaxy, and the columns 2 to 13 are the S\'ersic index in the 12 S-PLUS filters. A full version of this table, for all galaxies, is available as supporting information in the online version  available as supporting information in the online version.} \label{tab:Sersic_MegaMorph} 
\resizebox{\textwidth}{!}{
\begin{tabular}{lllllllllllllllll}

\hline
\hline
ID & $u$ & J0378 & J0395 & J0410 & J0430 & $g$ & J0515 & $r$ & J0660 & $i$ & J0861 & $z$ \\
\hline

1 & 1.36$\pm$0.02 & 1.23$\pm$0.02 & 1.17$\pm$0.02 & 1.12$\pm$0.01 & 1.07$\pm$0.01 & 0.97$\pm$0.02 & 0.94$\pm$0.01 & 0.95$\pm$0.01 & 0.99$\pm$0.01 & 1.10 $\pm$0.01 & 1.15 $\pm$0.01 & 1.13 $\pm$0.01\\
\hline

\end{tabular}
}
\end{table*}

\begin{table*}
\caption{Physical parameters of the galaxies. Column 1 is the ID of each galaxy, columns 2 and 3 are the right ascension and declination respectively. The stellar mass calculated with the colour $(g-i)$ presented in \citet{Taylor2011} is in column 4. The luminosity and sSFR are in the columns 5 and 6 respectively. A full version of this table, for all galaxies, is available as supporting information in the online version of this manuscript.} \label{tab:parameter}
    \centering
    \begin{tabular}{ccccccccc}

    \hline
    \hline
    ID & RA$^{\circ}$  & DEC$^{\circ}$  & $\log(M_{\star}/M_{\odot})$ & L$_{H\alpha}$ & $\log(sSFR/yr)$  \\
  & J2000 & J2000 &  & $10^{39}$ergs$^{-1}$ &  \\
    \hline

    1   & 159.67 & -28.57 & 10.01$\pm$0.05   & 64.48$\pm$7.34 & -10.01$\pm$0.06 \\
    \hline
    \end{tabular}
\end{table*}

\section{Simulated galaxies}
\label{appendix:simulation}. 

\begin{figure*}

\includegraphics[width=\textwidth]{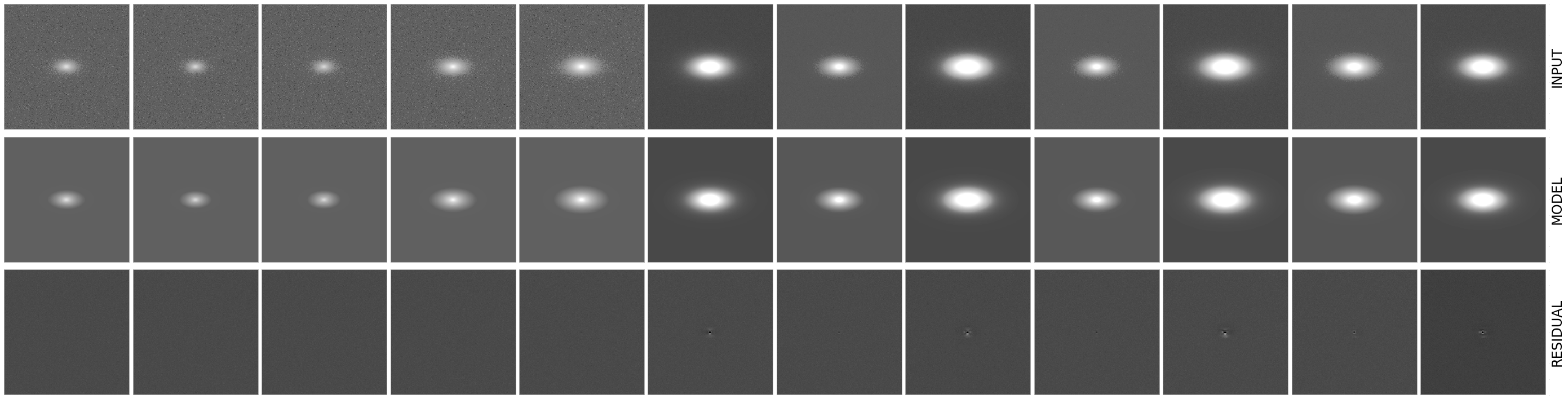}
    \caption{Simulated galaxy 1 (top panels), the \textsc{GALFITM} models (middle panels), and the residual image (observed minus modelled -- bottom panels).}
    \label{fig:G2_Q}
\end{figure*}

\begin{figure*}

\includegraphics[width=\textwidth]{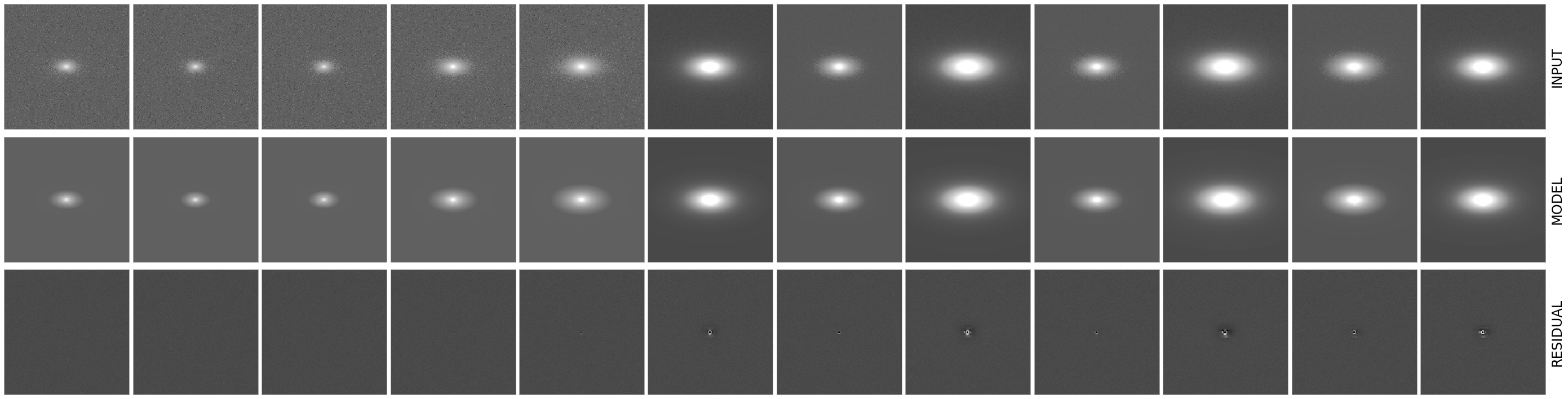}
    \caption{Simulated galaxy 2 (top panels), the \textsc{GALFITM} models (middle panels), and the residual image (observed minus modelled -- bottom panels)}
    \label{fig:G4_Q}
\end{figure*}

\begin{figure*}

\includegraphics[width=\textwidth]{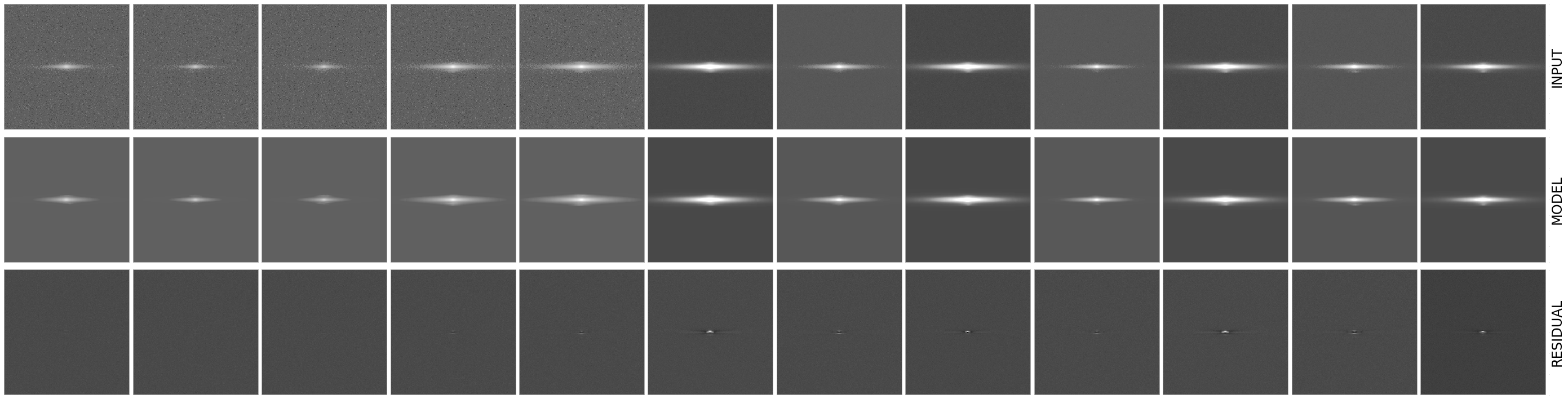}
    \caption{Simulated galaxy 3 (top panels), the \textsc{GALFITM} models (middle panels), and the residual image (observed minus modelled -- bottom panels).}
    \label{fig:G4_SF}
\end{figure*}

\begin{figure*}

\includegraphics[width=\textwidth]{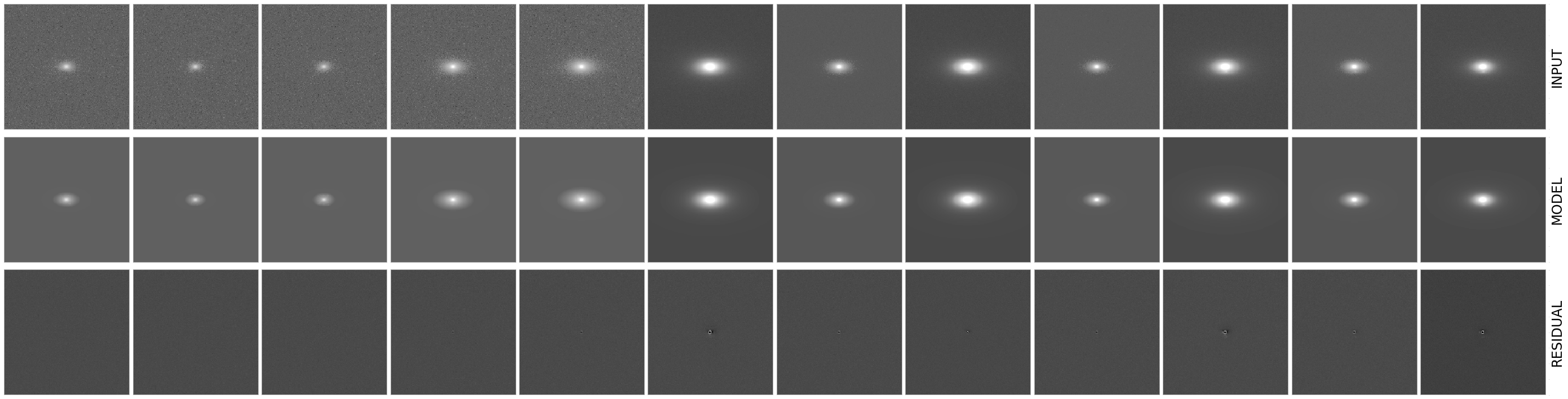}
    \caption{Simulated galaxy 4 (top panels), the \textsc{GALFITM} models (middle panels), and the residual image (observed minus modelled -- bottom panels).}
    \label{fig:G5_SF}
\end{figure*}

\begin{figure*}

\includegraphics[width=\textwidth]{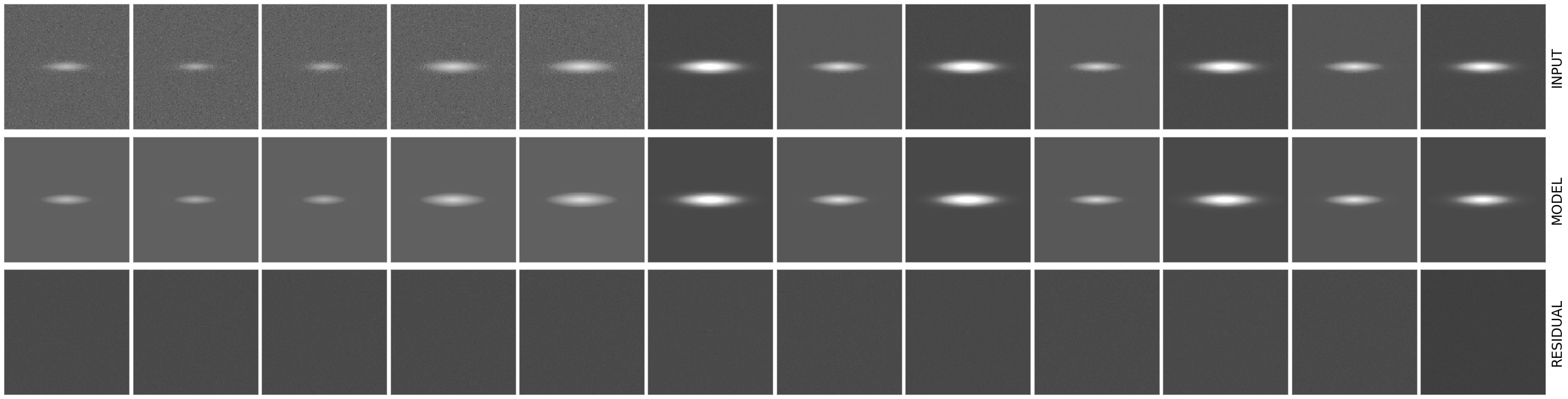}
    \caption{Simulated galaxy 5 (top panels), the \textsc{GALFITM} models (middle panels), and the residual image (observed minus modelled -- bottom panels).}
    \label{fig:G1_SF}
\end{figure*}

In order to prove the goodness of \textsc{GALFITM} on retrieving structural parameters and physical information, we generated a set of five simulated galaxies to be modelled by \textsc{GALFITM}. The simulated galaxies were generated in each of all SPLUS-filters, using the same range of observational parameters as in the observations (S/N, filters and background level). We use a star-forming SED (SF) and a quiescent SED (Q) to model a realistic wavelenght dependance of the flux. On this exercise, we fix the S\'ersic index over all wavelength, letting free the total flux. \textsc{GALFITM} allows us to recover the S\'ersic index and effective radius with an uncertainty $\sim$4 percent with respect to the value used in the construction of the simulated galaxy. In addition, we perform a linear regression, comparing the magnitudes of the simulated galaxies with respect to the magnitudes found by \textsc{GALFITM}. We find a coefficient of determination of $\sim$1. These results confirm that the parameters recovered by the \textsc{GALFITM} models are reliable. Table~\ref{tab:mag} lists the magnitudes in all 12 S-PLUS filters, $n_{r}$, $R_{e,r}$ and the SED used in the  five simulated galaxies (Input simulation) and those recovered by \textsc{GALFITM} (GALFITM output). Figures~\ref{fig:G2_Q}, ~\ref{fig:G4_Q}, ~\ref{fig:G4_SF}, ~\ref{fig:G5_SF} and ~\ref{fig:G1_SF}, show the 5 modelled galaxies. On these figures, top panels show the simulated galaxies, middle panels show the \textsc{GALFITM} models and bottom panels show the residuals, derived from the subtraction between the top and middle panels.

\begin{table*}
\caption{Magnitudes for 5 simulated galaxies (Input Model) and its magnitudes recovered by GALFITM (GALFITM output) }
\label{tab:mag} 

\resizebox{\textwidth}{!}{
\begin{tabular}{lllllllllllllllll}
Galaxy              & $u$     & J0378 & J0395 & J0410 & J0430  & $g$     & J0515 & $r$     & J0660 & $i$     & J0861 & z & $n_{r}$ & $R_{e,r}$ & SED    \\
& & & & & & & & & & & & & &(arcsec) & \\
\hline
\hline
1 Input simulation   & 13.75 & 13.53 & 13.34 & 13.08 & 12.59 & 12.17 & 11.90 & 11.52 & 11.45 & 11.20 & 11.04 & 11.00  & 2.00 & 10.00 & Q\\
1 GALFITM output & 13.74 & 13.52 & 13.33 & 13.07 & 12.58 & 12.16 & 11.89 & 11.50 & 11.44 & 11.19 & 11.03 & 11.00 & 1.98 & 10.02 \\
\hline
2 Input simulation   & 13.75 & 13.53 & 13.34 & 13.08 & 12.59 & 12.17 & 11.90 & 11.52 & 11.45 & 11.20 & 11.04 & 11.00 & 4.00 & 10.00 & Q\\
2 GALFITM output & 13.69 & 13.47 & 13.28 & 13.02 & 13.53 & 12.11 & 11.84 & 11.46 & 11.39 & 11.14 & 10.98 & 10.94 & 3.87 & 9.96\\
\hline
3 Input simulation   & 14.41 & 14.30 & 14.11 & 13.57 & 13.28 & 13.20 & 13.14 & 13.08 & 13.08 & 13.04 & 13.01 & 13.00 & 4.00 & 20.00 & SF\\
3 GALFITM output & 14.34 & 14.24 & 14.05 & 13.50 & 13.21 & 13.13 & 13.08 & 13.02 & 13.02 & 12.98 & 12.94 & 12.93 & 3.85 & 19.65 \\
\hline
4 Input simulation   & 14.41 & 14.30 & 14.11 & 13.57 & 13.28 & 13.20 & 13.14 & 13.08 & 13.08 & 13.04 & 13.01 & 13.00  & 5.00 & 10.00 & SF\\
4 GALFITM output & 14.33 & 14.22 & 14.03 & 13.48 & 13.19 & 13.11 & 13.05 & 13.00 & 13.00 & 12.96 & 12.92 & 12.92 & 4.82 & 9.68\\
\hline
5 Input simulation   & 14.41 & 14.30 & 14.11 & 13.57 & 13.28 & 13.20 & 13.14 & 13.08 & 13.08 & 13.04 & 13.01 & 13.00 & 1.00 & 20.00 & SF\\
5 GALFITM output & 14.42 & 14.31 & 14.11 & 13.56 & 13.27 & 13.19 & 13.13 & 13.08 & 13.07 & 13.04 & 13.00 & 12.99 & 0.99 & 20.01 \\
\end{tabular}
}
\end{table*}

%%%%%%%%%%%%%%%%%%%%%%%%%%%%%%%%%%%%%%%%%%%%%%%%%%

% Don't change these lines
\bsp	% typesetting comment
\label{lastpage}
\end{document}